\documentclass{article}
\usepackage{latexsym}
\usepackage{braket}
\usepackage{mathrsfs}
\usepackage{mathtools}
\usepackage{color}
\usepackage{xcolor}
\usepackage{subcaption}
\usepackage{enumitem}
\usepackage[toc,page]{appendix}
\usepackage{cite}
\usepackage[english]{babel}
\usepackage{babelbib}
\usepackage{multirow}
\usepackage{tabularx}
\usepackage{jcappub}
\usepackage[T1]{fontenc}
\usepackage{makecell}
\setcellgapes{4pt}
\usepackage[a4paper, total={6in, 8in}]{geometry}

\usepackage[thinlines]{easytable}

\usepackage{colortbl}
\def\bal#1\eal{\begin{align}#1\end{align}}
\usepackage{color,soul}
\newcommand{\bsub}{\begin{subequations}}
\newcommand{\esub}{\end{subequations}}
\def\bal#1\eal{\begin{align}#1\end{align}}

\newcommand{\gra}{{\alpha}} \newcommand{\grb}{{\beta}} \newcommand{\grg}{{\gamma}} \newcommand{\grd}{{\delta}}
\newcommand{\gre}{{\epsilon}} \newcommand{\grz}{{\zeta}} \newcommand{\grh}{{\eta}} \newcommand{\gru}{{\theta}}
\newcommand{\gri}{{\iota}} \newcommand{\grk}{{\kappa}} \newcommand{\grl}{{\lambda}} \newcommand{\grm}{{\mu}}
\newcommand{\grn}{{\nu}} \newcommand{\grj}{{\xi}}  \newcommand{\grp}{{\pi}}
\newcommand{\grr}{{\rho}} \newcommand{\grs}{{\sigma}} \newcommand{\grt}{{\tau}} 
\newcommand{\grf}{{\phi}}

  \newcommand{\grL}{{\Lambda}}

\title{\boldmath Dynamically equivalent $\mathbf{\grL CDM}$ equations with underlying Bianchi Type geometry}

\author[a]{T. Pailas,}
\author[a]{T. Christodoulakis,}

\affiliation[a]{ National and Kapodistrian University of Athens, \\Department of Nuclear and Particle Physics, \\Faculty of Physics, Athens 15784, Greece}

\emailAdd{teopailas879@hotmail.com}
\emailAdd{tchris@phys.uoa.gr}

\abstract{Solutions have been found for gravity coupled to electromagnetic field and a set of charged and uncharged perfect fluids for Bianchi Types $VI_{(-1)}$, $VIII$, $IX$. It has been assumed that the anisotropy is ``frozen'', $\grg_{\grm\grn}=\gra(t)^{2}m_{\grm\grn}$, where $\grg_{\grm\grn}$ and $m_{\grm\grn}$ are the spatial metric and some constant matrix respectively. This, according to previous works, results in the existence of a conformal Killing vector field proportional to the fluid velocity of the comoving matter, which guarantees the absence of parallax effects and the independence of the temperature (assuming black body spectrum) from the direction of observation. The electromagnetic field ``absorbs'' the ``frozen'' anisotropy and the remaining equations are dynamically equivalent with the equations of $\grL CDM$. There are solutions with flat, negative and positive effective spatial curvature corresponding to the three FLRW classes. Three equations of state for the charged perfect fluid were studied: non-relativistic $w=0$, relativistic $w=\frac{1}{3}$ and dark energy-like $w=-1$. For the first two cases, maximum values exist for the scale factor, in order for the weak energy conditions to be respected, which depend upon the geometric and charged fluid parameters. A minimum value for the scale factor exists (for the solutions to be valid) in all the cases and Types, indicating the absence of initial spacetime singularity (big bang). This minimum value depends upon the geometric and electromagnetic parameters. The number of essential constants in the final form of each metric is the minimum without loss of generality due to the use of the constant Automorphism's group. A known solution, with the anisotropy absorbed via one free scalar field is reproduced with our method and contains the minimum possible number of parameters.}

\begin{document}
\maketitle
\flushbottom

\section{Introduction}
\label{sec1}

The most recent data of Planck collaboration \cite{2018arXiv180706209P} indicate spatial homogeneity and isotropy of the CMB at large-scales. So far, the model used to describe the observations is the so called $\grL CDM$ with the underlying geometry of an FLRW metric. As it is known, the characteristic of an FLRW metric is that it is spatially homogeneous and isotropic, thus it seems the simplest possible candidate to describe the observations. A question arises: Does the observational data of the CMB uniquely fix the spacetime metric? 

J. Ehlers, P. Geren and R. K. Sachs have addressed to some extend, in the form of a theorem known as (EGS), the question of whether isotropy of CMB implies the isotropy of the spacetime metric \cite{doi:10.1063/1.1664720}. Generalizations of this theorem were presented in \cite{1999CQGra..16.3781C}. Further work in this direction can be found in \cite{1978MNRAS.184..439E, 1992PhRvD..46..578F, Stoeger:1994qs}. A set of spacetimes which maintain the homogeneity but lacks isotropy, come by the name Bianchi Types. The anisotropy of these spacetimes is in general dynamical, which would cause large-scale anisotropies in CMB far from the observed values, \cite{Barrow:1985tda, Hervik2011, Thorsrud2012}. As it was proven in \cite{1988JMP....29.2064H}, a spacetime which admits a conformal Killing vector field proportional to the velocity vector field of the comoving matter, leads to a parallax-free universe. Furthermore, its existence guarantees that the temperature T of the radiation field (assuming black body spectrum) will be independent of the direction of observation \cite{1992GReGr..24..121O}. Note that, in order for the Bianchi Types to admit such a conformal Killing vector field, their anisotropy must be ``frozen'' and not dynamical i.e. $\grg_{\grm\grn}=\gra(t)^{2}m_{\grm\grn}$, where $\grg_{\grm\grn}$ and $m_{\grm\grn}$ are the spatial metric and some constant matrix respectively.

The next question would concern the nature of the fields capable to ``absorb'' this ``frozen'' anisotropy. In \cite{Carneiro:2001fz} the authors studied the case of Bianchi Type III with the existence of one free scalar field. They manage to find a possible solution. The distance-redshift relations and the estimation of the cosmological parameters in this particular model were studied in \cite{2013JCAP...03..033M}. Models which are shear-free but anisotropic were studied in \cite{Mimoso1993, Coley1994, McManus1994, 2011GReGr..43.3253E}. The investigation of spatial distribution of supernovae in terms of Bianchi models with FLRW behavior was addressed in \cite{Koivisto:2010dr}. An interesting and extensive work was carried out by Mikjel Thorsrud in \cite{Thorsrud2018} were he used a set of n independent p-form gauge fields in order to ``absorb'' the anisotropy. He has proven that the only Bianchi Type whose the anisotropy can be absorbed by only one free scalar field with positive energy density, is the Locally Rotationally Symmetric (LRS) Bianchi Type III, previously studied in \cite{Carneiro:2001fz}. Finally, cosmological perturbation theory in anisotropic backgrounds was employed in the following works, \cite{1986PhRvD343570T, Pereira2007, EmirGmrko2007,  Kofman:2011tr, Franco2017}.

The existence of electromagnetic fields in the Bianchi spacetimes have been studied by many authors though the years. Solutions with large scale magnetic fields have been found in \cite{1970ApJ...160..147H}. In the work \cite{Sagnotti:1981wq} the authors have reduced the propagation problem of electromagnetic waves for the case of Bianchi Type I, to the integration of a second order differential equation. M.S. Madsen studied the behavior of scalar electrodynamics under symmetry breaking by the Higgs mechanism in the class of Bianchi Type I spacetime \cite{Madsen:1994qm}. M. Wollensak has shown that solutions analogous to plane wave solutions in flat spacetime, must obey two transversal conditions when there are at least two scale factors \cite{Wollensak:1998mx}. Exact solutions of a mass-less dilaton field interacting with an electromagnetic field have been found for Bianchi Types I and III as well as the Kantowski-Sachs, in \cite{Banerjee:1999sw}. The search for singularities in spaces conformal related to Bianchi Types, under the presence of nonlinear electrodynamics have been studied in \cite{GarciaSalcedo:2004nc}. The author Kei Yamamoto showed, by using dynamical system analysis, that a family of plane-wave solutions of the Einstein-Maxwell's equations are the stable attractor for expanding universes in the case of Bianchi Class B spacetimes \cite{Yamamoto:2011xy}. The quantum analysis of a Bianchi III LRS geometry coupled to a source free electromagnetic field was presented in \cite{Karagiorgos:2017nta}. Most recently, some of the solutions found in \cite{Pailas:2019abb} have the property of being electromagnetic pp-wave spacetimes and also belong to a special case of a 3D ``Bianchi'' Type (i.e. spacetimes which admit a two dimensional group with simply transitive action on two dimensional surfaces).

This work aims to ``absorb'' the ``frozen'' anisotropy by use of electromagnetic field. As it was proven in \cite{Carneiro:2001fz}, \cite{Thorsrud2018}, the free electromagnetic field cannot succeed. To this end, we will assume the existence of a charged fluid carrying a four-current density, interacting with the electromagnetic field. This combination of matter sources, is more likely to have appeared during the Big Bang nucleosynthesis and the Photon epoch where the temperature of the universe remained too high in order for neutral atoms to appear. Thus, we consider that this choice is somehow physically justified. Furthermore, the group of constants Automorphisms will be used at the spatial metric. The importance of this lies in the following; once the Automorphisms are used the metric is cleared of all the non-essential constants (non-essential in the sense that there is coordinate transformation capable to absorb them). The number of the remaining essential constants will be the minimum and this is advantageous when the solutions are compared with observational data, since we can identify exactly which of the constants possess a specific physical meaning, for instance, the only constant in the Schwarzschild metric is related to the mass of the compact object. Thus, we achieve the simplest possible form of the solutions, without loss of generality. As we shall see, solutions can be found only for the cases of Bianchi Type $VI_{(-1)}$, $VIII$, $IX$. We reproduce the solution of Bianchi Type III in the presence of a free scalar field with our method in order to make that clear.

The structure of the paper is organized as follows: in section \ref{sec2} the mathematical preliminaries concerning with the basic equations used and the group of Automorphisms are presented. The basic assumption for the spacetime to admit a conformal Killing vector field is introduced in section \ref{sec3} as well as the application of the group of constant Automorphisms. In section \ref{sec4} the reproduction of the solution involving one free scalar field is presented. The section \ref{sec5} is dedicated to the ``absorption'' of the ``frozen'' anisotropy via electromagnetic field in interaction with a charged fluid. A discussion of the overall results and things to come can be found in section \ref{sec6}. Finally, an Appendix is also included. 

\section{Mathematical Preliminaries}
\label{sec2}

\subsection{Bianchi Types}

Let us start with the line element of a four dimensional manifold M
\begin{align}
{ds^{2}_{(4)}}=\left(-N^{2}(t,x)+N_{i}(t,x)N^{i}(t,x)\right)dt^{2}+2N_{i}(t,x)dx^{i}dt+\gamma_{ij}(t,x)dx^{i}dx^{j},\label{e14}
\end{align}
where $i,j$ are coordinate indices running from $1$ to $3$. The $3+1$ analysis has been used, where $N(t,x)$ is the lapse, $N_{i}(t,x)$ the shift and $\grg_{ij}(t,x)$ the metric of the spatial hypersurfaces $t=\text{constant}$. Note also that $N_{i}(t,x)N^{i}(t,x)=\grg^{ij}(t,x)N_{i}(t,x)N_{j}(t,x)$ where $\grg^{ij}(t,x)$ the inverse of $\grg_{ij}(t,x)$. It is well known that there are coordinates $(\tilde{t},\tilde{x}^{i})$,  usually called Gaussian normal coordinates \cite{2007gr.qc.....3035G}, such that the line element \eqref{e14} acquires the form
\begin{align}
{ds^{2}_{(4)}}=-d\tilde{t}^{2}+\tilde{\gamma}_{ij}(\tilde{t},\tilde{x})d\tilde{x}^{i}d\tilde{x}^{j},\label{eqs14}
\end{align}
or equivalently $\tilde{N}(\tilde{t},\tilde{x})=1$, $\tilde{N}_{i}(\tilde{t},\tilde{x})=0$. For simplicity, we omit the symbol ``tilde'' from the coordinates and the spatial metric, thus
\begin{align}
{ds^{2}_{(4)}}=-dt^{2}+\gamma_{ij}(t,x)dx^{i}dx^{j}.\label{eqs15}
\end{align}
Let us now restrict our attention to spatially (simply) homogeneous spacetimes. Recall that a spacetime is called spatially (simply) homogeneous when a three dimensional isometry group G acts simply transitively on the three dimensional hypersurfaces $t=constant$. When the action is simply transitive, there exists an invariant basis of one-forms ${\grs^{\gra}}$ satisfying the curl relations \cite{1975hrc..book.....R}, \cite{Nakahara:2003nw}
\begin{align}
d\grs^{\grl}=-\frac{1}{2}C^{\grl}_{\gra\grb}\grs^{\gra}\wedge\grs^{\grb}\Leftrightarrow \partial_{i}\grs^{\grl}_{j}-\partial_{j}\grs^{\grl}_{i}=-C^{\grl}_{\gra\grb}\grs^{\gra}_{i}\grs^{\grb}_{j},\label{e15}
\end{align}
such that
\begin{align}
{\cal{L}}_{\grj_{\gra}}\grs^{\grb}=0,\label{e145}
\end{align}
where the Greek indices $(\gra,\grb,...)$ run from 1 to 3 and count the different triads, $C^{\grl}_{\gra\grb}$ are the structure constants of the Lie algebra of the isometry group and have the property $C^{\grl}_{\gra\grb}=-C^{\grl}_{\grb\gra}$, $\{\grj_{\gra}\}$ is the set of the Killing fields and ${\cal{L}}_{\grj_{\gra}}$ the Lie derivative along them. The corresponding Killing fields for each Bianchi Type can be found in \cite{1975hrc..book.....R}, \cite{2013arXiv1304.7894T}. The line element \eqref{eqs15} can then be written in the manifestly homogeneous form
\begin{align}
{ds^{2}_{(4)}}=-dt^{2}+\gamma_{\gra\grb}(t)\sigma^{\gra}_{i}(x)\sigma^{\grb}_{j}(x)dx^{i}dx^{j},\label{e16}
\end{align}
or equivalently
\begin{align}
{ds^{2}_{(4)}}=-dt^{2}+\gamma_{\gra\grb}(t)\sigma^{\gra}(x)\sigma^{\grb}(x),\label{e16879}
\end{align}
where $\grs^{a}(x)=\grs^{\gra}_{i}(x)dx^{i}$. The case where spatial homogeneity was imposed on the original line element \eqref{e14} can be found in the Appendix A.

\subsection{Group of Constant Automorphisms}

The group of coordinate transformations that preserve the hypersurface's manifest homogeneity and, as a side effect, also generate symmetries of the Einstein's equations are called ``rigid'' symmetries \cite{1993CQGra..10.1607C}. For transformations of the form
\begin{align}
t\mapsto\tilde{t}=t, \hspace{2.0cm}\\
x^{i}\mapsto{\tilde{x}}^{i}=h^{i}(x^{l}), \hspace{0.2cm}x^{i}=f^{i}({\tilde{x}}^{l}),
\end{align}
the restrictions on the functions $f^{i}$, in order for the manifest homogeneity to be preserved, are summarized as follows
\begin{align}
\sigma^{\alpha}_{i}(x^{l})\frac{\partial x^{i}}{\partial {\tilde{x}}^m}&={\Lambda^{\alpha}}_{\beta}\sigma^{\beta}_{m}({\tilde{x}}^{l}).\label{es31}
\end{align}
The relations \eqref{es31} must be regarded as the definition of the matrix ${\Lambda^{\alpha}}_{\beta}$. A generalization of this for the case of time dependent matrix  ${\Lambda^{\alpha}}_{\beta}(t)$, as well as the implications of the above equation \eqref{es31} can be found in Appendix A. The line element \eqref{e16} can then be written as
\begin{align}
{ds^{2}_{(4)}}=-d{\tilde{t}}^{2}+{\tilde{\gamma}}_{\alpha\beta}\left(\tilde{t}\right)\sigma^{\alpha}_{i}({\tilde{x}}^{l})\sigma^{\beta}_{j}({\tilde{x}}^{l})d{\tilde{x}}^{i}d{\tilde{x}}^{j},
\end{align}
with the abbreviation
\begin{align}
\tilde{\grg}_{\gra\grb}(\tilde{t})&=\grg_{\grm\grn}(\tilde{t}){\grL^{\grm}}_{\gra}{\grL^{\grn}}_{\grb}.\label{es50}
\end{align}
The existence of local solutions to the equations \eqref{es31} is guaranteed by the Frobenious theorem if the following necessary and sufficient condition holds (for more on this subject take a look at the Appendix A):
\begin{align}
{\Lambda^{\alpha}}_{\mu}C^{\mu}_{\beta\nu}&=C^{\alpha}_{\mu\sigma}{\Lambda^{\mu}}_{\beta}{\Lambda^{\sigma}}_{\nu}.\label{es34}
\end{align}
The solutions of \eqref{es34} form the so called constant Automorphisms group. Given the structure constants of the group, the matrix ${\Lambda^{\alpha}}_{\mu}$ is determined. The number of the independent, non-zero components of $\Lambda$ provides the dimension of the group. How can someone use the freedom provided by the matrix ${\grL^{\gra}}_{\grb}$? As we can see from $\eqref{es50}$, the spatial metric is time-dependent while the matrix ${\grL^{\gra}}_{\grb}$ is constant. Thus, a direct application of this equation does not have much to offer, since we will not be able to simplify the spatial metric. However, one could consider the equation \eqref{es50} as Lie point transformations \cite{stephani_1990}, \cite{olver_1995} of the dependent variable $\grg_{\grm\grn}$ since, as we have already stated, the group of Automorphisms generate symmetries of Einstein's equations. The non-zero elements of $\grL$ will be the parameters of the Lie symmetry group. This is of great importance since it allows for the reduction of order of the Einstein's equations and in most of the cases the entire solution space to be found without loss of generality. A series of papers aligned in this direction are \cite{Christodoulakis:2004yx, Christodoulakis:2006vi, Terzis:2008ev}. Another way to use the group of constant Automorphisms is in the special case that concerns the present work, where
\begin{align}
\grg_{\grm\grn}(t)=a^{2}(t)m_{\grm\grn},\label{er56}
\end{align}
with $m_{\grm\grn}$ some constant symmetric matrix. Thus, the equivalent of the equation $\eqref{es50}$ would be
\begin{align}
\tilde{m}_{\grm\grn}=m_{\grm\grn}{\grL^{\grm}}_{\gra}{\grL^{\grn}}_{\grb}.\label{er60}
\end{align}
In that case, the matrix ${\grL^{\grm}}_{\gra}$ could be directly used for the simplification of the matrix $m_{\grm\grn}$ and equivalently for the spatial metric without loss of generality. The form \eqref{er56} is going to be justified in the upcoming sections. Why to use the freedom provided by ${\grL^{\grm}}_{\gra}$? From a mathematical perspective, the expressions for the equations as well as the objects involved will be greatly simplified. Furthermore, this is reflected to the fact that the remaining arbitrary non-zero constants of $\tilde{m}_{\grm\grn}$ will be essential, or in other words, there will be no further coordinate transformations that can absorb them. Thus, any possible physical or geometrical meaning of the solutions, will be attached to these constants (like, e.g. the integration constant appearing in the Schwarzschild metric, corresponding to the mass of a point-like source, with proper unit conventions). As we shall see in the forthcoming sections, the constants will be related to the effective spatial curvature of the spacetime.

\subsection{System of equations}

Let us write the equations of the system that we are going to study in this paper and explaining each one of them. Apart from minor differences, we follow the conventions of \cite{2007gr.qc.....3035G}. We assume coordinates $(t,x^{i})$, where the Latin characters $(i,j,l,...)$ take values from 1 to 3, thus every object will be a function of these coordinates. The system consists of gravity coupled to a set of fluids. For the purposes of this work, the total energy momentum tensor $T_{(tot)}^{\grm\grn}$ splits into three parts: a part corresponding to electromagnetically uncharged matter $T_{(u)}^{\grm\grn}$, a charged one $T_{(c)}^{\grm\grn}$ and the electromagnetic part $T_{(em)}^{\grm\grn}$, where the Greek indices $\grm,\grn$ run from 1 to 4.

\textbf{Einstein's Field Equations (EFE)}
\begin{align}
&R^{(3)}+K^{2}-K_{ij}K^{ij}=2\grk \grr^{(tot)},\\
&D_{i}K-D_{j}{K_{i}}^{j}=\grk q_{i}^{(tot)},
\end{align}
\begin{align}
\partial_{t}K_{ij}-{\cal{L}}_{\vec{N}}K_{ij}=N R^{(3)}_{ij}-N\left(2{K_{i}}^{l}K_{lj}-K K_{ij}\right)-D_{j}D_{i}N-\grk N\left[\grp_{ij}^{(tot)}+\frac{1}{2}\left(\grr^{(tot)}-P^{(tot)}\right)\grg_{ij}\right].
\end{align}
We have used the $3+1$ analysis and the adapted coordinate system, with the introduction of the lapse $(N)$ and shift $(N_{i})$. The symbol $D_{i}$ is used to denote the covariant derivative related to the metric $\grg_{ij}$ of the three dimensional hypersurfaces $t=constant$. The objects $R^{(3)}_{ij}, R^{(3)},K_{ij},K$, correspond to the Ricci tensor, Ricci scalar, extrinsic curvature, and the trace of the extrinsic curvature of the hypersurfaces. The Lie derivative along the shift vector is represented by ${\cal{L}}_{\vec{N}}$ and will be used wherever is needed. Note that $K_{ij}$ is expressed in terms of $\grg_{ij}$, $N$ and $N_{i}$ as follows
\begin{align}
K_{ij}=-\frac{1}{2N}\left(\partial_{t}\grg_{ij}-D_{j}N_{i}-D_{i}N_{j}\right).
\end{align}
When it comes to the total energy momentum tensor, the fluid decomposition has been employed along the vector field $n=\left(\frac{1}{N},-\frac{N^{i}}{N}\right)$ normal to the hypersurfaces. The quantities $\grr^{(tot)}$,
$P^{(tot)}$, $q_{i}^{(tot)}$, $\grp_{ij}^{(tot)}$ correspond to the density, isotropic pressure, flux and the traceless part of the anisotropic pressure tensor. Finally, $\grk$ is the coupling constant $\grk=\frac{8\grp G}{c^{4}}$ where G the Newton's gravitational constant and c the speed of light.

Now we proceed with the conservation of the total energy momentum tensor which will provide us with the equations of motion for the fluid parts.\\

\textbf{Uncharged Matter Field Equations (UMFE)}
\begin{align}
\nabla_{\grn}T^{\grm\grn}_{(u)}=0\Rightarrow\nonumber
\end{align}
\begin{align}
&\partial_{t}\grr_{(u)}-{\cal{L}}_{\vec{N}}\grr_{(u)}-\left(\grr_{(u)}+P_{(u)}\right)N K-D_{i}\left(N q^{i}_{(u)}\right)-q^{i}_{(u)}D_{i}N-N \grp^{ij}_{(u)}K_{ij}=0,\\
&\partial_{t}q_{i}^{(u)}-{\cal{L}}_{\vec{N}}q_{i}^{(u)}-\left(\grr^{(u)}+P^{(u)}\right)D_{i}N-N D_{i}P^{(u)}-N K q^{(u)}_{i}-D_{j}\left(N \grp^{(u)j}_{i}\right)=0.
\end{align}

\textbf{Charged Matter Field Equations (CMFE)}
\begin{align}
\nabla_{\grn}T^{\grm\grn}_{(c)}+\nabla_{\grn}T^{\grm\grn}_{(em)}=0\Rightarrow
\nabla_{\grn}T^{\grm\grn}_{(c)}-F^{\grm\grs}J_{\grs}=0\Rightarrow\nonumber
\end{align}
\begin{align}
&\partial_{t}\grr_{(c)}-{\cal{L}}_{\vec{N}}\grr_{(c)}-\left(\grr_{(c)}+P_{(c)}\right)N K-D_{i}\left(N q^{i}_{(c)}\right)-q^{i}_{(c)}D_{i}N-N \grp^{ij}_{(c)}K_{ij}=N E^{i}J_{i},\\
&\partial_{t}q_{i}^{(c)}-{\cal{L}}_{\vec{N}}q_{i}^{(c)}-\left(\grr^{(c)}+P^{(c)}\right)D_{i}N-N D_{i}P^{(c)}-N K q^{(c)}_{i}-D_{j}\left(N \grp^{(c)j}_{i}\right)=-N\left(\grr_{(e)}E_{i}+B_{ij}J^{j}\right),
\end{align}
where $\nabla_{\grm}$ is the covariant derivative related to the four dimensional spacetime, $F^{\grm\grs}$ the Faraday tensor and $J_{\grs}$ the four-current, which has as time component the charge density $\grr_{(e)}$ and as spatial component the three-current $J_{i}$. We have also use the relation $\nabla_{\grn}T^{\grm\grn}_{(em)}=-F^{\grm\grs}J_{\grs}$ which holds modulo the Maxwell's equations. Note that $E_{i}, B_{ij}$ are the corresponding electric and magnetic fields, with $B_{ij}=-B_{ji}$. The definition of $B_{ij}$ is $B_{ij}=F_{ij}$. When $i,j$ run from $1$ to $3$, there is the correspondence $B_{ij}=\gre_{ijk}B^{k}$ where $\gre_{ijk}$ is the totally antisymmetric symbol. Thus, $B^{\grk}=\frac{1}{2}\gre^{kij}B_{ij}=\frac{1}{2}\gre^{kij}F_{ij}$. The definition of $B_{ij}$ is valid for any dimension therefore we choose to use this insted of $B^{\grk}$. The next step are the Maxwell's equations.\\

\textbf{Maxwell's Field Equations (MFE)}
\begin{align}
&D_{i}E^{i}=\grm_{0}\grr_{(e)},\,\,\,\partial_{t}E^{i}-{\cal{L}}_{\vec{N}}E^{i}-D_{j}\left(N B^{ij}\right)-N K E^{i}+\grm_{0}N J^{i}=0,\\
&D_{[l}B_{ij]}=0,\,\,\,\,\,\,\partial_{t}B_{ij}-{\cal{L}}_{\vec{N}}B_{ij}+2D_{[i}(NE_{j]})=0.
\end{align}
The only comments here are that the symbol $[]$ stands for total anti-symmetrization of the indices enclosed and $\grm_{0}$ is the magnetic permeability of vacuum. Finally, due to (MFE) the conservation of charge follows\\

\textbf{Charge Conservation Field Equations (CCFE)}
\begin{align}
\nabla_{\grm}J^{\grm}=0\Rightarrow \partial_{t}\grr_{(e)}-{\cal{L}}_{\vec{N}}\grr_{(e)}-N K \grr_{(e)}+D_{i}\left(N J^{i}\right)=0.
\end{align}

\subsection{The system of equations for Bianchi Types}
We present the equations of the previous section under the assumption of spatial homogeneity provided by the existence of a simply transitive group acting on the hypersurfaces $t=\text{constant}$.\\

\textbf{(EFE)}
\begin{align}
&R^{(3)}+K^{2}-K_{\gra\grb}K^{\gra\grb}=2\grk\grr^{(tot)},\label{e17}\\
&{K_{\gra}}^{\grb}C^{\gra}_{\grb \grl}+{K_{\grl}}^{\grb}C^{\gra}_{\grb\gra}=\grk q_{\grl}^{(tot)},\label{e18}
\end{align}
\begin{align}
\dot{K}_{\gra\grb}=N R_{\gra\grb}^{(3)}-N\left(2{K_{\gra}}^{\grl}K_{\grl\grb}-K K_{\gra\grb}\right)-N^{\gre}\left(K_{\gra \grl}C^{\grl}_{\gre\grb}+K_{\grb \grl}C^{\grl}_{\gre\gra}\right)-\grk N\left(\grp_{\gra\grb}^{(tot)}+\frac{\grr^{(tot)}-P^{(tot)}}{2}\grg_{\gra\grb}\right),\label{e19}
\end{align}
where the extrinsic and the Ricci curvature are given by the following expressions
\begin{align}
K_{\gra\grb}=-\frac{1}{2N}\left(\dot{\grg}_{\gra\grb}+N^{\grl}C^{\gre}_{\grl\grb}\grg_{\gre\gra}+N^{\grl}C^{\gre}_{\grl\gra}\grg_{\gre\grb}\right),
\end{align}
\begin{align}
R_{\grm\grn}=-\frac{1}{2}C^{\gra}_{\grb \grm}\left(C^{\grb}_{\gra \grn}+\grg^{\grb \gre}\grg_{\gra \grz}C^{\grz}_{\gre\grn}\right)+\frac{1}{4}\grg_{\grm \gra}\grg_{\grn \grb}\grg^{\grz \grt}\grg^{\gre \gru}C^{\gra}_{\grz \gre}C^{\grb}_{\grt \gru}-\frac{1}{2}C^{\grb}_{\grt\grb}\grg^{\grt \gra}\left(C^{\grz}_{\gra \grm}\grg_{\grn \grz}+C^{\grz}_{\gra \grn}\grg_{\grm \grz}\right).\label{e21}
\end{align}

\textbf{(UMFE)}
\begin{align}
&\dot{\grr}_{(u)}-\left(\grr_{(u)}+P_{(u)}\right) N K +N q^{\grm}_{(u)}C^{\gra}_{\grm\gra}-N \grp^{\gra\grm}_{(u)}K_{\gra\grm}=0,\\
&\dot{q}_{\grm}^{(u)}+N^{\grb}q^{(u)}_{\gra}C^{\gra}_{\grb\grm}-N K q^{(u)}_{\grm}+N\left(C^{\gra}_{\grb \gra}\grp_{\grm}^{(u)\grb}+\grp_{\gra}^{(u)\grb}C^{\gra}_{\grb\grm}\right)=0.
\end{align}

\textbf{(CMFE)}
\begin{align}
&\dot{\grr}_{(c)}-\left(\grr_{(c)}+P_{(c)}\right) N K +N q^{\grm}_{(c)}C^{\gra}_{\grm \gra}-N \grp^{\gra\grm}_{(c)}K_{\gra\grm}=N E^{\gra}J_{\gra},\\
&\dot{q}_{\grm}^{(c)}+N^{\grb}q^{(c)}_{\gra}C^{\gra}_{\grb\grm}-N K q^{(c)}_{\grm}+N\left(C^{\gra}_{\grb\gra}\grp_{\grm}^{(c)\grb}+\grp_{\gra}^{(c)\grb}C^{\gra}_{\grb\grm}\right)=-N\left(\grr_{(e)}E_{\grm}+B_{\grm \gra}J^{\gra}\right).
\end{align}

\textbf{(MFE)}
\begin{align}
&C^{\gra}_{\grm\gra}E^{\grm}=-\grm_{0}\grr_{(e)},\,\,\,\dot{E}^{\grm}+E^{\grb}N^{\gra}C^{\grm}_{\grb \gra}+N\left(B^{\grm \grl}C^{\gra}_{\grl \gra}+\frac{1}{2}C^{\grm}_{\grl \gra}B^{\grl \gra}\right)-N K E^{\grm}+\grm_{0}N J^{\grm}=0,\\
&B_{\grm[\gra}C^{\grm}_{\grb\grl]}=0,\,\,\,\,\,\,\dot{B}_{\grm\grn}+N^{\gra}\left(B_{\grm \grl}C^{\grl}_{\gra\grn}+B_{\grl\grn}C^{\grl}_{\gra\grm}\right)-NE_{\grl}C^{\grl}_{\grm\grn}=0.
\end{align}

\textbf{(CCFE)}
\begin{align}
\dot{\grr}_{(e)}-N K \grr_{(e)}-N C^{\gra}_{\grm\gra}J^{\grm}=0.
\end{align}
Every quantity of the above equations is only t dependent, thus the $(\cdot)$ denotes derivative with respect to t. For the electric and magnetic field, as well as the current density, we have assumed that $E_{i}(t,x)=E_{\gra}(t)\grs^{\gra}_{i}(x)$, $B_{ij}(t,x)=B_{\gra\grm}(t)\grs^{\gra}_{i}(x)\grs^{\grm}_{j}(x)$ and $ J_{i}(t,x)=J_{\gra}(t)\grs^{\gra}_{i}(x)$. The reason why this is an assumption is explained in the Appendix D.

\section{Primary assumption and use of Automorphisms}
\label{sec3}

As we have already pointed out in the introduction, it has been proven that a spacetime which admits a conformal Killing vector field proportional to the vector field of the comoving radiation fluid, will be parallax-free and the temperature (assuming black body spectrum) will not depend on the direction of observation. For the Bianchi Types to admit such a conformal Killing vector field the following needs to be assumed
\begin{align}
\grg_{\grm\grn}=a(t)^{2}m_{\grm\grn},\label{e51}
\end{align}
where $m_{\grm\grn}$ is a $3\times3$ constant symmetric matrix. Note that, the indices $\grm,\grn$ are triad indices and run through $1$ to $3$. The inverse is given by
\begin{align}
\grg^{\grm\grn}=\frac{1}{a(t)^{2}}m^{\grm\grn},\,\, \text{such that}\,\,\grg^{\grl\grm}\grg_{\grm\grn}=\grd^{\grl}_{\grn}\Rightarrow m^{\grl\grm}m_{\grm\grn}=\grd^{\grl}_{\grn}.
\end{align}
Under this assumption and with the previously justified choices $N_{i}(t,x)=0$, $N(t,x)=1$, the line element of the spacetime becomes
\begin{align}
ds^{2}_{(4)}=-dt^{2}+\gra(t)^{2}m_{\grm\grn}\grs^{\grm}(x)\grs^{\grn}(x).\label{ormetr}
\end{align}
With the phrase ``frozen'' anisotropy we refer to the existence of only one scale factor $\gra(t)$, and thus there is coordinate $\tilde{t}$ such that the line element can be written in the ``conformal'' time gauge as
\begin{align}
ds^{2}_{(4)}=\tilde{\gra}(\tilde{t})\left[-d\tilde{t}^{2}+m_{\grm\grn}\grs^{\grm}(x)\grs^{\grn}(x)\right],
\end{align}
The conformal Killing vector field in the original and/or final coordinates is given by
\begin{align}
\grj_{(c)}=\gra(t)n,\,\text{or},\, \grj_{(c)}=\tilde{a}(\tilde{t})\tilde{n}
\end{align}
where $n=\partial_{t}$ or $\tilde{n}=\frac{1}{\tilde{\gra}(\tilde{t})}\partial_{\tilde{t}}$ is the unit normal to the surfaces $t=\text{constant}$ and which in our analysis corresponds to the comoving vector field. It is easy to verify that for each Bianchi Type, the following holds
\begin{align}
{\cal{L}}_{\grj_{(c)}}g_{IJ}=2\dot{\gra}(t)g_{IJ},
\end{align}
where $g_{IJ}$ the spacetime metric associated with the line element\eqref{ormetr}, $(I,J=1,2,3,4)$. The equations now become

\textbf{(EFE)}
\begin{align}
&H^{2}=\grk\frac{\grr^{(tot)}}{3}-\frac{\tilde{R}}{6}\frac{1}{a^{2}},\label{e54}\\
&\dot{H}+H^{2}=-\grk\frac{1}{6}\left(\grr^{(tot)}+3P^{(tot)}\right),\label{e55}
\end{align}
\begin{align}
&q^{(tot)}_{\grm}=0,\label{e56}\\
&\grp^{(tot)}_{\grm\grn}=\frac{1}{\grk}\left(\tilde{R}_{\grm\grn}-\frac{1}{3}\tilde{R}m_{\grm\grn}\right),\label{e57}
\end{align}
where the time dependence has been suppressed and $H=\frac{\dot{a}}{a}$ is the Hubble function. We provide more details of how these equations came up in the Appendix B. Note that $\tilde{R}_{\grm\grn}$ is constant and is given by \eqref{e21} where $\grg_{\grm\grn}$ is replaced by $m_{\grm\grn}$, hence $\tilde{R}=m^{\grm\grn}\tilde{R}_{\grm\grn}$. By the redefinition $\tilde{R}=6\,k$ (where k is the representative of the spatial curvature in FLRW metric) the equations \eqref{e54}, \eqref{e55} are the same in form as the ones where the underlying geometry is that of FLRW metric. For this to happen, the total flux of the fluids should be zero, \eqref{e56}, while the anisotropic pressure should be given by \eqref{e57}. The ``frozen'' anisotropy of Bianchi Types is absorbed, once the equations \eqref{e56}, \eqref{e57} are satisfied. Let us also provide the rest of the equations.\\

\textbf{(UMFE)}
\begin{align}
&\dot{\grr}_{(u)}+3\left(\grr_{(u)}+P_{(u)}\right) H +q^{\grm}_{(u)}C^{\gra}_{\grm\gra}=0,\\
&\dot{q}_{\grm}^{(u)}+3 q^{(u)}_{\grm}H+\left(C^{\gra}_{\grb \gra}\grp_{\grm}^{(u)\grb}+\grp_{\gra}^{(u)\grb}C^{\gra}_{\grb\grm}\right)=0.
\end{align}

\textbf{(CMFE)}
\begin{align}
&\dot{\grr}_{(c)}+3\left(\grr_{(c)}+P_{(c)}\right) H +q^{\grm}_{(c)}C^{\gra}_{\grm \gra}=E^{\gra}J_{\gra},\\
&\dot{q}_{\grm}^{(c)}+3 q^{(c)}_{\grm}H+\left(C^{\gra}_{\grb\gra}\grp_{\grm}^{(c)\grb}+\grp_{\gra}^{(c)\grb}C^{\gra}_{\grb\grm}\right)=-\left(\grr_{(e)}E_{\grm}+B_{\grm \gra}J^{\gra}\right).
\end{align}

\textbf{(MFE)}
\begin{align}
&C^{\gra}_{\grm\gra}E^{\grm}=-\grm_{0}\grr_{(e)},\,\,\,\dot{E}^{\grm}+3 E^{\grm}H+\left(B^{\grm \grl}C^{\gra}_{\grl \gra}+\frac{1}{2}C^{\grm}_{\grl \gra}B^{\grl \gra}\right)+\grm_{0} J^{\grm}=0,\\
&B_{\grm[\gra}C^{\grm}_{\grb\grl]}=0,\,\,\,\dot{B}_{\grm\grn}-E_{\grl}C^{\grl}_{\grm\grn}=0.
\end{align}

\textbf{(CCFE)}
\begin{align}
\dot{\grr}_{(e)}+3 \grr_{(e)}H-C^{\gra}_{\grm\gra}J^{\grm}=0.
\end{align}

Before we proceed in the search of fluids which will ``absorb'' this ``frozen'' anisotropy, let us see how the Automorphisms will prove useful. The equations \eqref{e51} and \eqref{es50} provides us with the relation
\begin{align}
\tilde{m}_{\grm\grn}=m_{\gra\grb}{\grL^{\gra}}_{\grm}{\grL^{\grb}}_{\grn},
\end{align}
thus we can use the constant Automorphisms in order to simplify as much as possible the matrix $m_{\gra\grb}$. This is of great importance, since the remaining components of $m_{\gra\grb}$ will correspond to essential constants, in the sense that there will be no coordinate transformation able to absorb them. The number of constants that remained in the metric after the use of Automorphisms is the minimum. For completeness, we provide in a table the matrices, ${\grL^{\gra}}_{\grm}$, $m_{\gra\grb}$ for each one of the Bianchi Types. The structure constants for the Bianchi Types that we use in this work can be found in \cite{1975hrc..book.....R}, \cite{doi:10.1063/1.523441}, \cite{2013arXiv1304.7894T}. In order to be compatible with our conventions, an overall minus sign is needed $C^{\gra}_{\grb\grm}\rightarrow-C^{\gra}_{\grb\grm}$. We present only the non-zero independent structure constants for each Type in the table below, as we have used them in order to obtain the result.

\begin{center}
\begin{tabular}{ |c|c|c| } 
\hline
\textbf{Bianchi Type, Structure Constants} & \textbf{Automorphism} & \textbf{Metric} \\
\hline
I & $\begin{pmatrix}
e^{b_{1}} & b_{2} & b_{3}\\
b_{4} & e^{b_{5}} & b_{6}\\
b_{7} & b_{8} & e^{b_{9}}
\end{pmatrix}$ & $\begin{pmatrix}
1 & 0 & 0\\
0 & 1 & 0\\
0 & 0 & 1
\end{pmatrix}$\\
\hline
II, $C^{1}_{23}=-1$ & $\begin{pmatrix}
e^{b_{5}+b_{6}}-b_{3}b_{4} & b_{1} & b_{2}\\
0 & e^{b_{5}} & b_{3}\\
0 & b_{4} & e^{b_{6}}
\end{pmatrix}$ & $\begin{pmatrix}
1 & 0 & 0\\
0 & 1 & 0\\
0 & 0 & m_{1}
\end{pmatrix}$\\
\hline
III, $C^{1}_{13}=-1$ & $\begin{pmatrix}
e^{b_{1}} & 0 & b_{2}\\
0 & e^{b_{3}} & b_{4}\\
0 & 0 & 1
\end{pmatrix}$ & $\begin{pmatrix}
1 & m_{1} & 0\\
m_{1} & 1 & 0\\
0 & 0 & m_{2}
\end{pmatrix}$\\
\hline
IV, $C^{1}_{13}=-1$, $C^{1}_{23}=-1$, $C^{2}_{23}=-1$ & $\begin{pmatrix}
e^{b_{1}} & b_{2} & b_{3}\\
0 & e^{b_{1}} & b_{4}\\
0 & 0 & 1
\end{pmatrix}$ & $\begin{pmatrix}
1 & 0 & 0\\
0 & m_{1} & 0\\
0 & 0 & m_{2}
\end{pmatrix}$\\
\hline
V, $C^{1}_{13}=-1$, $C^{2}_{23}=-1$ & $\begin{pmatrix}
e^{b_{1}} & b_{2} & b_{3}\\
b_{4} & e^{b_{5}} & b_{6}\\
0 & 0 & 1
\end{pmatrix}$ & $\begin{pmatrix}
1 & 0 & 0\\
0 & 1 & 0\\
0 & 0 & m_{1}
\end{pmatrix}$\\
\hline
$VI_{(h)}$, $h\neq\{0,1\}$, $C^{1}_{13}=-1$, $C^{2}_{23}=-h$ & $\begin{pmatrix}
e^{b_{1}} & 0 & b_{2}\\
0 & e^{b_{3}} & b_{4}\\
0 & 0 & 1
\end{pmatrix}$ & $\begin{pmatrix}
1 & m_{1} & 0\\
m_{1} & 1 & 0\\
0 & 0 & m_{2}
\end{pmatrix}$\\
\hline
$VII_{(h)}$, $h\geq{0}$, $C^{1}_{13}=-h$, $C^{2}_{13}=1$, $C^{1}_{23}=-1$, $C^{2}_{23}=-h$ & $\begin{pmatrix}
e^{b_{1}} & -b_{2} & b_{3}\\
b_{2} & e^{b_{1}} & b_{4}\\
0 & 0 & 1
\end{pmatrix}$ & $\begin{pmatrix}
1 & 0 & 0\\
0 & m_{1} & 0\\
0 & 0 & m_{2}
\end{pmatrix}$\\
\hline
VIII, $C^{1}_{23}=1$, $C^{2}_{13}=1$, $C^{3}_{12}=-1$ & ${\grL_{(1)\gra}^{\grm}}{\grL_{(2)\grb}^{\gra}}{\grL_{(3)\grn}^{\grb}}$ & $\begin{pmatrix}
m_{1} & 0 & 0\\
0 & m_{2} & 0\\
0 & 0 & m_{3}
\end{pmatrix}$\\
\hline
IX, $C^{1}_{23}=-1$, $C^{2}_{13}=1$, $C^{3}_{12}=-1$ & ${\grL_{(4)\gra}^{\grm}}{\grL_{(5)\grb}^{\gra}}{\grL_{(6)\grn}^{\grb}}$ & $\begin{pmatrix}
m_{1} & 0 & 0\\
0 & m_{2} & 0\\
0 & 0 & m_{3}
\end{pmatrix}$\\
\hline
\end{tabular}
\captionof{table}{This table contains the structure constants, the constant Automorphisms matrices and the irreducible form of the target space spatial metric for each Bianchi Type.}
\end{center}
Note that the constant $m_{1}$ appearing in the metrics of Types III, VI should be bounded in the domain (-1,1). In all the other cases, the constants $m_{1}$, $m_{2}$, $m_{3}$, should be positive in order for $m_{\gra\grb}$ to be positive definite and the spacetime metric to have a Lorentzian signature. Also, for the Types VIII, IX the Automorphism matrices are given below.
\begin{align}
&{\grL_{(1)\gra}^{\grm}}=\begin{pmatrix}
Cosh (b_{1}) & Sinh(b_{1}) & 0\\
Sinh(b_{1}) & Cosh(b_{1}) & 0\\
0 & 0 & 1
\end{pmatrix}, 
{\grL_{(2)\grb}^{\gra}}=\begin{pmatrix}
Cosh (b_{2}) & 0 & Sinh(b_{2})\\
0 & 1 & 0\\
Sinh(b_{2}) & 0 & Cosh (b_{2})
\end{pmatrix},
{\grL_{(3)\grn}^{\grb}}=\begin{pmatrix}
1 & 0 & 0\\
0 & Cos(b_{3}) & -Sin(b_{3})\\
0 & Sin(b_{3}) & Cos(b_{3})
\end{pmatrix},\nonumber\\
&{\grL_{(4)\gra}^{\grm}}=\begin{pmatrix}
Cos (b_{1}) & -Sin(b_{1}) & 0\\
Sin(b_{1}) & Cos(b_{1}) & 0\\
0 & 0 & 1
\end{pmatrix}, 
{\grL_{(5)\grb}^{\gra}}=\begin{pmatrix}
Cos (b_{2}) & 0 & -Sin(b_{2})\\
0 & 1 & 0\\
Sin(b_{2}) & 0 & Cos (b_{2})
\end{pmatrix},
{\grL_{(6)\grn}^{\grb}}=\begin{pmatrix}
1 & 0 & 0\\
0 & Cos(b_{3}) & -Sin(b_{3})\\
0 & Sin(b_{3}) & Cos(b_{3})
\end{pmatrix}.\nonumber
\end{align}
Another way to list the different Bianchi Types, is based on the Behr decomposition in which the structure constants are decomposed as follows
\begin{align}
C^{k}_{ij}=\gre_{ijl}\grh^{lk}+a_{l}\left(\grd^{k}_{i}\grd^{l}_{j}-\grd^{k}_{j}\grd^{l}_{i}\right),
\end{align}
where $a_{l}$, $\grh^{lk}$ are given by
\begin{align}
&a_{i}=-\frac{1}{2}C^{j}_{ij},\\
&\grh^{mk}=C^{(k}_{ij}\gre^{m)ij}.
\end{align}
More information, as well as the expression of the Ricci tensor and scalar of the hypersurfaces in terms of $a_{l}$, $\grh^{lk}$ can be found in \cite{Hervik}. In this way, one can use the form of the Ricci scalar given in this reference and verify the validity of the results presented in the following table.

The next step is to calculate the Ricci tensor, the Ricci scalar and then the traceless anisotropic pressure tensor as it is given from \eqref{e57}. We provide a table with $\grp_{\gra\grb}^{(tot)}$ and the Ricci scalar in order to comment about whether the hypersurface has positive, negative or zero curvature.
\begin{center}
\begin{tabular}{ |c|c|c| } 
\hline
\textbf{Bianchi Type} & $\grp_{\gra\grb}^{(tot)}$ & $\tilde{R}$ \\
\hline
I & $\begin{pmatrix}
0 & 0 & 0\\
0 & 0 & 0\\
0 & 0 & 0
\end{pmatrix}$ & 0 \\
\hline
II & $\frac{1}{\grk}\begin{pmatrix}
\frac{2}{3 m_{1}} & 0 & 0\\
0 & -\frac{1}{3 m_{1}} & 0\\
0 & 0 & -\frac{1}{3}
\end{pmatrix}$ & $-\frac{1}{2 m_{1}}<0$\\
\hline
III & $\frac{1}{\grk}\frac{1}{3\left(1-m_{1}^{2}\right)}\begin{pmatrix}
\frac{-1+3 m_{1}^{2}}{m_{2}} & \frac{2 m_{1}}{m_{2}} & 0\\
\frac{2 m_{1}}{m_{2}} & \frac{2}{m_{2}} & 0\\
0 & 0 & -1
\end{pmatrix}$ & $ -\frac{4-3 m_{1}^{2}}{2 m_{2}\left(1-m_{1}^{2}\right)}<0$ \\
\hline
IV & $\frac{1}{\grk}\frac{1}{3 m_{1}}\begin{pmatrix}
\frac{2}{m_{2}} & -\frac{3 m_{1}}{m_{2}} & 0\\
-\frac{3 m_{1}}{m_{2}} & -\frac{m_{1}}{m_{2}} & 0\\
0 & 0 & -1
\end{pmatrix}$ & $-\frac{1+12 m_{1}}{2 m_{1} m_{2}}<0$ \\
\hline
V & $\begin{pmatrix}
0 & 0 & 0\\
0 & 0 & 0\\
0 & 0 & 0
\end{pmatrix}$ & $-\frac{6}{m_{1}}<0$ \\
\hline
VI, $h\neq{0,1}$ & $\frac{1}{\grk}\begin{pmatrix}
\frac{\left(-1+h\right)\left(1+2 h-3 m_{1}^{2}\right)}{3\left(1-m_{1}^{2}\right)m_{2}} & \frac{2\left(-1+h\right)^{2}m_{1}}{3\left(1-m_{1}^{2}\right)m_{2}} & 0\\
\frac{2\left(-1+h\right)^{2}m_{1}}{3\left(1-m_{1}^{2}\right)m_{2}} & \frac{f}{3\left(1-m_{1}^{2}\right)m_{2}} & 0\\
0 & 0 & \frac{-(-1+h)^{2}}{3\left(1-m_{1}^{2}\right)}
\end{pmatrix}$ & $-\frac{4(1+h+h^{2})-3(1+h)^{2}m_{1}^{2}}{2\left(1-m_{1}^{2}\right)m_{2}}<0$\\
\hline
VII, $h\geq{0}$ & $\frac{1}{\grk}\begin{pmatrix}
\frac{2-m_{1}\left(1+m_{1}\right)}{3 m_{1} m_{2}} & \frac{h\left(-1+m_{1}\right)}{m_{2}} & 0\\
\frac{h\left(-1+m_{1}\right)}{m_{2}} & \frac{-1+m_{1}(2m_{1}-1)}{3 m_{2}} & 0\\
0 & 0 & -\frac{\left(-1+m_{1}\right)^{2}}{3 m_{1}}
\end{pmatrix}$ & $-\frac{1+m_{1}\left(-2+12 h^{2}+m_{1}\right)}{2 m_{1} m_{2}}<0$\\
\hline
VIII & $\frac{1}{\grk}\begin{pmatrix}
\frac{f_{1}}{3 m_{2} m_{3}} & 0 & 0\\
0 & \frac{f_{2}}{3 m_{1} m_{3}} & 0\\
0 & 0 & \frac{f_{3}}{3 m_{1} m_{2}}
\end{pmatrix}$ & $-\frac{m_{1}\left[m_{1}+2( m_{2}+m_{3})\right]+\left(m_{2}-m_{3}\right)^{2}}{2 m_{1} m_{2} m_{3}}<0$\\
\hline
IX & $\frac{1}{\grk}\begin{pmatrix}
\frac{f_{4}}{3 m_{2} m_{3}} & 0 & 0\\
0 & \frac{f_{5}}{3 m_{1} m_{3}} & 0\\
0 & 0 & \frac{f_{6}}{3 m_{1} m_{2}}
\end{pmatrix}$ & $-\frac{m_{1}\left[m_{1}-2( m_{2}+m_{3})\right]+\left(m_{2}-m_{3}\right)^{2}}{2 m_{1} m_{2} m_{3}}$\\
\hline
\end{tabular}
\captionof{table}{The traceless anisotropic pressure tensor and the Ricci scalar $\tilde{R}$ is presented in this table. Also, due to the value and the sign of $\tilde{R}$, the curvature of the spatial hypersurface is characterized as positive, negative or zero.}
\end{center}

The following abbreviations were used, for \textbf{Type VI} $f=(-1+h)\left[-2+h(-1+3m_{1}^{2})\right]$, for \textbf{Type VIII} $f_{1}=m_{1}\left(2m_{1}+m_{2}+m_{3}\right)-\left(m_{2}-m_{3}\right)^{2}$, $f_{2}=m_{2}\left(2m_{2}+m_{1}-m_{3}\right)-\left(m_{1}+m_{3}\right)^{2}$, $f_{3}=m_{3}\left(2m_{3}+m_{1}-m_{2}\right)-\left(m_{1}+m_{2}\right)^{2}$ and for \textbf{Type IX} $f_{4}=m_{1}\left(2m_{1}-m_{2}-m_{3}\right)-\left(m_{2}-m_{3}\right)^{2}$, $f_{5}=m_{2}\left(2m_{2}-m_{1}-m_{3}\right)-\left(m_{1}-m_{3}\right)^{2}$, $f_{6}=m_{3}\left(2m_{3}-m_{1}-m_{2}\right)-\left(m_{1}-m_{2}\right)^{2}$. 

For Type IX we haven't used an inequality symbol and the reason is that in this Type all the three cases can be achieved. Specifically, 
\begin{equation}
\tilde{R}_{(IX)}=
\begin{cases}
  \begin{aligned}
  \geq{0}, & & m_{2}+m_{3}-2\sqrt{m_{2}m_{3}} &\leq m_{1} \leq m_{2}+m_{3}+2\sqrt{m_{2}m_{3}}, \\
  <0  , & & & \text{otherwise.}\\
  \end{aligned}
\end{cases}
\end{equation}
It is in the limit of zero anisotropy that the Bianchi Type IX has a positive Ricci scalar. To make this clear, let as assume a small perturbation
\begin{align}
&m_{i}=1+\tilde{\gre}_{i}, \hspace{0.2cm} i=1,2,3,\\
&\tilde{\gre}_{i}<<1,
\end{align}
where $1$ is the background value (leading to the closed FLRW). Under this  assumption, the last entry of the Table 2 becomes
\begin{align}
-\frac{m_{1}\left[m_{1}-2( m_{2}+m_{3})\right]+\left(m_{2}-m_{3}\right)^{2}}{2 m_{1} m_{2} m_{3}}=\frac{3}{2}-\frac{1}{2}\left(\tilde{\gre}_{1}+\tilde{\gre}_{2}+\tilde{\gre}_{3}\right)+\mathcal{O}\left(\tilde{\gre}^{2}\right).
\end{align}
Thus, the background value of the Ricci scalar is positive as it should. The corrections on the other hand, can assume any value. For more information on this subject, we provide the following works \cite{Pontzen:2010eg}, \cite{1991PhRvD..44.2356K}.
 
The Types I and V correspond to flat and open FLRW spaces respectively, thus will not concern us further. The closed FLRW is provided from Type IX when $m_{1}=m_{2}=m_{3}$, therefore this case will not concern us either. These cases will not concern us because their traceless anisotropic pressure tensor is zero, thus correspond to spatially isotropic and homogeneous spacetimes (FLRW), while we are interesting in the cases where anisotropy is present. Now, we study the fields which will ``absorb'' the anisotropy represented by $\grp_{\gra\grb}^{(tot)}$.

\section{``Absorption'' via free scalar field}
\label{sec4}

In this section, there is no electromagnetic field, neither an electrically charged fluid nor four current, thus the equations \textbf{(CMFE)}, \textbf{(MFE)} and \textbf{(CCFE)} are identically satisfied. When it comes to the uncharged fluid, we assume that it consists of a set of perfect fluids (dust, radiation, cosmological constant), and one free scalar field. Furthermore, those fluids are non-interacting. Taking all that into account we may write
\begin{align}
&\grr^{(tot)}=\grr^{(d)}+\grr^{(r)}+\grr^{(\grL)}+\grr^{(\grf)},\\
&P^{(tot)}=P^{(d)}+P^{(r)}+P^{(\grL)}+P^{(\grf)},\\
&q^{(tot)}_{\grm}=q^{(\grf)}_{\grm},\label{e71}\\
&\grp^{(tot)}_{\grm\grn}=\grp^{(\grf)}_{\grm\grn},
\end{align}
where $(d)$ corresponds to dust, $(r)$ to radiation, $(\grL)$ to cosmological constant and $(\grf)$ to scalar field. Thus, the total flux and the traceless anisotropic pressure tensor are equated to those of the free scalar field. The four dimensional energy momentum tensor and the corresponding $(3+1)$ fluid quantities for the scalar field are given below 
\begin{align}
T_{\grm\grn}=M\left(\partial_{\grm}\grf\partial_{\grn}\grf-\frac{1}{2}g_{\grm\grn}\partial_{\grs}\grf\partial^{\grs}\grf\right),
\end{align}
where $M$ is some constant. Note that, the $\grm,\grn$ are coordinate-basis indices and run from $1$ to $4$.
\begin{align}
\grr^{(\grf)}&=\frac{M}{2}\left[(\partial_{t}\grf)^{2}+\partial_{i}\grf\partial^{i}\grf\right],\,\,P^{(\grf)}=\frac{M}{2}\left[(\partial_{t}\grf)^{2}-\frac{1}{3}\partial_{i}\grf\partial^{i}\grf\right],\\
q^{(\grf)}_{i}&=M\partial_{t}\grf\partial_{i}\grf,\label{e76}\\
\grp^{(\grf)}_{ij}&=M\left(\partial_{i}\grf\partial_{j}\grf-\frac{1}{3}\grg_{ij}\partial_{l}\grf\partial^{l}\grf\right).\label{e77}
\end{align}
The equations \eqref{e56}, \eqref{e71}, \eqref{e76},  combined lead to either $\partial_{t}\grf=0$ or $\partial_{i}\grf=0$. If the later holds, from \eqref{e77} we get zero traceless anisotropic pressure tensor. Thus, the only choice for our purpose is $\partial_{t}\grf=0$. Furthermore, the term $\partial_{i}\grf$ is necessarily expressed in the Bianchi basis, $\partial_{i}\grf(t,x)=\grf_{\gra}\grs^{a}_{i}(x)$, where $\grf_{\gra}$ some constant array, satisfying the following condition
\begin{align}
\grf_{\gra}C^{\gra}_{\grb\grl}=0.
\end{align} 
The indices $\gra,\grb,\grl$ run from $1$ to $3$. A rigorous proof of this statement can be found in the Appendix C. Taken that into account the fluid quantities become\\
\begin{align}
&\grr^{(\grf)}=\frac{M}{2a^{2}}\grf_{\gra}\grf^{\gra},\,\,P^{(\grf)}=-\frac{M}{6 a^{2}}\grf_{\gra}\grf^{\gra},\\
&q^{(\grf)}_{\gra}=0,\,\,\grp^{(\grf)}_{\gra\grb}=M\left(\grf_{\gra}\grf_{\grb}-\frac{1}{3}m_{\gra\grb}\grf_{\grl}\grf^{\grl}\right),
\end{align}
where $\grf_{\gra}\grf^{\gra}=\grf_{\gra}m^{\gra\grb}\grf_{\grb}$.

By use of the equations of state for the three perfect fluids, and the previously said about the scalar field, the equations \textbf{(EFE)} and \textbf{(UMFE)} become\\

\textbf{(EFE)}
\begin{align}
&H^{2}=\frac{\grk}{3}\left(\grr^{(d)}+\grr^{(r)}+\grr^{(\grL)}\right)-\frac{1}{6}\frac{\left(\tilde{R}-\grk M \grf_{\gra}\grf^{\gra}\right)}{a^{2}},\\
&\dot{H}+H^{2}=-\frac{\grk}{6}\left(\grr^{(d)}+2\grr^{(r)}-2\grr^{(\grL)}\right),\\
&M\left(\grf_{\gra}\grf_{\grb}-\frac{1}{3}m_{\gra\grb}\grf_{\grl}\grf^{\grl}\right)=\frac{1}{\grk}\left(\tilde{R}_{\gra\grb}-\frac{1}{3}\tilde{R}m_{\gra\grb}\right).\label{e85}
\end{align}

\textbf{(UMFE)}
\begin{align}
&\dot{\grr}^{(d)}+3\grr^{(d)}H=0,\label{e86}\\
&\dot{\grr}^{(r)}+4\grr^{(r)}H=0,\label{e87}\\
&\dot{\grr}^{(\grL)}=0.\label{e88}
\end{align}

The \textbf{(UMFE)} for the scalar field were identically satisfied.
Once a solution is given for \eqref{e85}, the equations are identical in the form with those of $\mathbf{\grL CDM}$ with an effective hypersurface curvature given by $k_{(eff)}=\frac{1}{6}\left(\tilde{R}-\grk M \grf_{\gra}\grf^{\gra}\right)$. We will not extend further since already has been found in \cite{Carneiro:2001fz}, \cite{Thorsrud2018}, that the only case for which \eqref{e85} admits a solution and the field has a positive energy density is Type III. We present the solution by use of our method, cleared from any non-essential constants and without loss of generality. 
\begin{align}
&ds_{(4)}^{2}=-dt^{2}+a(t)^{2}\left(m_{2}dx^{2}+e^{-2x}dy^{2}+dz^{2}\right),\,\,\grf=\pm \frac{z}{\sqrt{\grk m_{2}M}},\\
&H^{2}=\frac{\grk}{3}\left(\frac{\grr^{(d)}_{0}}{a(t)^{3}}+\frac{\grr^{(r)}_{0}}{a(t)^{4}}+\grr^{(\grL)}_{0}\right)+\frac{1}{2m_{2}a(t)^{2}},
\end{align} 
where the well known solutions of the equations \eqref{e86}, \eqref{e87} and \eqref{e88} were used
\begin{align}
\grr^{(d)}=\frac{\grr^{(d)}_{0}}{a(t)^{3}},
\grr^{(r)}=\frac{\grr^{(r)}_{0}}{a(t)^{4}},
\grr^{(\grL)}=\grr^{(\grL)}_{0},
\end{align}
where $\grr^{(d)}_{0},\grr^{(r)}_{0},\grr^{(\grL)}_{0}$ some constants. The constant $m_{1}$ was equated to zero in order for a solution to exist. Finally, the effective curvature is $k_{(eff)}=-\frac{1}{2m_{2}}$, which means that it corresponds to an effectively open universe. The parameter $m_{2}$ is related to the curvature of the universe and is the only one related to the geometry. There is also the possibility to normalize the constant $M$ such that $M=\frac{1}{\grk m_{2}}$ and
\begin{align}
\grf=\pm z.
\end{align}

\section{``Absorption'' via electromagnetic field}
\label{sec5}

In this case, we consider the existence of a charge fluid which interacts with the electromagnetic field, in addition to the usual non-interacting perfect fluids (dust, radiation, cosmological constant). The charge fluid will also be considered to be a perfect fluid with an equation of state $P^{(c)}=w \grr^{(c)}$. That having been said, we have
\begin{align}
&\grr^{(tot)}=\grr^{(d)}+\grr^{(r)}+\grr^{(\grL)}+\grr^{(c)}+\grr^{(em)},\\
&P^{(tot)}=P^{(d)}+P^{(r)}+P^{(\grL)}+P^{(c)}+P^{(em)},\\
&q^{(tot)}_{\grm}=q^{(em)}_{\grm},\,\,\grp^{(tot)}_{\grm\grn}=\grp^{(em)}_{\grm\grn},
\end{align}

In this section, the indices $\grm,\grn,\grl$ are triad indices running from $1$ to $3$. Alongside with the assumption that we made in the mathematical preliminaries about the electric and magnetic fields in Bianchi Types, the fluid quantities for the electromagnetic field are

\begin{align}
&\grr^{(em)}=\frac{1}{2\grm_{0}}\left(E_{\grm}E^{\grm}+\frac{1}{2}B_{\grm\grn}B^{\grm\grn}\right),\,\,P^{(em)}=\frac{1}{6\grm_{0}}\left(E_{\grm}E^{\grm}+\frac{1}{2}B_{\grm\grn}B^{\grm\grn}\right),\\
&q^{(em)}_{\grm}=-\frac{1}{\grm_{0}}B_{\grm\grn}E^{\grn},\,\,\grp_{\grm\grn}^{(em)}=\frac{1}{\grm_{0}}\left(B_{\grm\grl}{B_{\grn}}^{\grl}-\frac{a^{2}}{3}B_{\grl\grs}B^{\grl\grs}m_{\grm\grn}-E_{\grm}E_{\grn}+\frac{a^{2}}{3}E_{\grl}E^{\grl}m_{\grm\grn}\right).
\end{align}

The electromagnetic field has an equation of state of the form $P^{(em)}=\frac{1}{3}\grr^{(em)}$. Note also that the inner products are calculated with $\grg_{\grm\grn}$. Let us write the equations to be solved.\\

\textbf{(EFE)}
\begin{align}
&H^{2}=\frac{\grk}{3}\left(\grr^{(d)}+\grr^{(r)}+\grr^{(\grL)}\right)-\frac{\tilde{R}}{6}\frac{1}{a^{2}}+\frac{\grk}{3}\left(\grr^{(c)}+\grr^{(em)}\right)\label{e99}\\
&\dot{H}+H^{2}=-\frac{\grk}{6}\left(\grr^{(d)}+2\grr^{(r)}-2\grr^{(\grL)}\right)-\frac{\grk}{6}\left[\left(1+3 w\right)\grr^{(c)}+2\grr^{(em)}\right],\label{e100}\\
&-\frac{1}{\grm_{0}}B_{\grm\grn}E^{\grn}=0,\label{e1001}\\
&\frac{1}{\grm_{0}}\left(B_{\grm\grl}{B_{\grn}}^{\grl}-\frac{a^{2}}{3}B_{\grl\grs}B^{\grl\grs}m_{\grm\grn}-E_{\grm}E_{\grn}+\frac{a^{2}}{3}E_{\grl}E^{\grl}m_{\grm\grn}\right)=\frac{1}{\grk}\left(\tilde{R}_{\grm\grn}-\frac{1}{3}\tilde{R}m_{\grm\grn}\right).\label{e1002}
\end{align}

\textbf{(UMFE)}
\begin{align}
\dot{\grr}^{(d)}+3\grr^{(d)}H=0,\,\,\dot{\grr}^{(r)}+4\grr^{(r)}H=0,\,\,\dot{\grr}^{(\grL)}=0.
\end{align}

\textbf{(CMFE)}
\begin{align}
\dot{\grr}^{(c)}+3(1+w)\grr^{(c)}H-E^{\grm}J_{\grm}=0,\,\,\grr_{(e)}E_{\grm}+B_{\grm\grn}J^{\grn}=0.
\end{align}

\textbf{(MFE)}
\begin{align}
&C^{\gra}_{\grm\gra}E^{\grm}=-\grm_{0}\grr_{(e)},\,\,\,\dot{E}^{\grm}+3 E^{\grm}H+\left(B^{\grm \grl}C^{\gra}_{\grl \gra}+\frac{1}{2}C^{\grm}_{\grl \gra}B^{\grl \gra}\right)+\grm_{0} J^{\grm}=0,\label{GMAL}\\
&B_{\grm[\gra}C^{\grm}_{\grb\grl]}=0,\,\,\,\dot{B}_{\grm\grn}-E_{\grl}C^{\grl}_{\grm\grn}=0.
\end{align}

\textbf{(CCFE)}
\begin{align}
\dot{\grr}_{(e)}+3 \grr_{(e)}H-C^{\gra}_{\grm\gra}J^{\grm}=0.
\end{align}

As we can see, the equations \eqref{e99}, \eqref{e100} have a contribution from the electromagnetic field and the charged fluid. It is not obvious as in the case of a free scalar field that those two components will contribute only to the spatial curvature. In the subsections to follow, we present the solutions found. 

\subsection{Type VI$\mathbf{_{(-1)}}$ or A$_\mathbf{{3,4}}$ or E$\mathbf{(1,1)}$}

In this section we will present, to some extent, how we have found the solutions, but in the forthcoming ones only the results will be presented. All the solutions were found with the \textit{Mathematica} \textcopyright\, software. Let us start by providing the following objects $E_{\grm}$, $B_{\grm\grn}$, $J_{\grm}$:
\begin{align}
&E_{\grm}=\left(E_{1}(t), E_{2}(t), E_{3}(t)\right),\label{enm1}\\
&B_{\grm\grn}=\begin{pmatrix}
0 & B_{1}(t) & -B_{2}(t)\\
-B_{1}(t) & 0 & B_{3}(t)\\
B_{2}(t) & -B_{3}(t) & 0
\end{pmatrix},\label{enm2}\\
&J_{\grm}=\left(J_{1}(t), J_{2}(t), J_{3}(t)\right).\label{enm3}
\end{align}
By use of the structure constants for this Type, we found out the following results:

\begin{align}
&\dot{B}_{\grm\grn}(t)-E_{\grl}(t)C^{\grl}_{\grm\grn}=0\Rightarrow\nonumber\\
&\dot{B}_{1}(t)=0, \label{l5}\\
&E_{1}(t)-\dot{B}_{2}(t)=0,\label{l6}\\
&E_{2}(t)-\dot{B}_{3}(t)=0.\label{l7}
\end{align}
From \eqref{l5} it follows that $B_{1}(t)=\grb$. The solutions to the equations \eqref{e1001}, \eqref{e1002} result
\begin{align}
&E_{1}(t)=0, E_{2}(t)=0, B_{2}(t)=0, B_{3}(t)=0,\\
&E_{3}(t)=\frac{\sqrt{-\grk m_{2} \grb^{2}+2 \grm_{0}a(t)^{2}}}{{\sqrt{\grk}a(t)}}, m_{1}=0.
\end{align}
Next we write the equations \eqref{GMAL} as
\begin{align}
&\grr_{(e)}(t)=0,\\
&J_{1}(t)=0,\\
&J_{2}(t)=0,\\
&J_{3}(t)+\frac{2\dot{\gra}(t)}{\sqrt{\grk}\sqrt{-\grk m_{2}\grb^{2}+2\grm_{0}\gra(t)^{2}}}=0,
\end{align}
the last of which implies $J_{3}(t)=-\frac{2\dot{\gra}(t)}{\sqrt{\grk}\sqrt{-\grk m_{2}\grb^{2}+2\grm_{0}\gra(t)^{2}}}$. Altogether we acquire
\begin{align}
&E_{\grm}=\left(0, 0, \frac{\sqrt{-\grk m_{2} \grb^{2}+2\grm_{0} \gra^{2}(t)}}{\sqrt{\grk}\gra(t)}\right),\label{enm8}\\
&B_{\grm\grn}=\begin{pmatrix}
0 & \grb & 0\\
-\grb & 0 & 0\\
0 & 0 & 0
\end{pmatrix},\label{enm9}\\
&J_{\grm}=\left(0, 0, -\frac{2\dot{\gra}(t)}{\sqrt{\grk}\sqrt{-\grk m_{2}\grb^{2}+2\grm_{0}\gra(t)^{2}}}\right).\label{enm10}
\end{align}
In order to express the line element and the Faraday tensor in the coordinate basis, let us recall the one-form basis components for this Type\cite{1975hrc..book.....R}
\begin{align}
\grs^{\grl}_{i}=\begin{pmatrix}
e^{-z} & 0 & 0\\
0 & e^{z} & 0\\
0 & 0 & 1
\end{pmatrix},
\end{align}
and find out the expressions for $E_{i}(t,x)=E_{\gra}(t)\grs^{\gra}_{i}(x)$, $B_{ij}(t,x)=B_{\gra\grm}(t)\grs^{\gra}_{i}(x)\grs^{\grm}_{j}(x)$ and 
$ J_{i}(t,x)=J_{\gra}(t)\grs^{\gra}_{i}(x)$. The Faraday tensor $\mathbf{F}$ and the four-current $\mathbf{J}$ are then given by $\mathbf{F}=-E_{i}dt\wedge dx^{i}+B_{ij}dx^{i}\wedge dx^{j}$, $\mathbf{J}=-\grr_{(e)}dt+J_{i}dx^{i}$, where $j>i, i,j=1,2,3$. Note that $x^{i}=\left(x,y,z\right)$. Finally, let us provide the expressions for the line element, the Faraday tensor and the four current
\begin{align}
&ds_{(4)}^{2}=-dt^{2}+a(t)^{2}\left(e^{-2z}dx^{2}+e^{2z}dy^{2}+m_{2}dz^{2}\right),\\
&\mathbf{F}=-\frac{\sqrt{-\grk m_{2} \grb^{2}+2 \grm_{0}a(t)^{2}}}{{\sqrt{\grk}a(t)}}dt\wedge dz+\grb dx\wedge dy, \\
&\mathbf{J}=-\frac{2\dot{a}(t)}{\sqrt{\grk}\sqrt{-\grk m_{2} \grb^{2}+2\grm_{0}a^{2}(t)}}dz,
\end{align}
where $\wedge$ stands for the wedge product $dt\wedge dz=dt \otimes dz-dz\otimes dt$. The original parameter $m_{1}$ is equal to zero and thus only $m_{2}$ remains. This metric admits only the three original Killing vectors fields of the Bianchi Type $VI_{(h)}$.

In order for the four current as well as the Faraday tensor to be real, the scale factor has to be bounded from below.
\begin{align}
-\grk m_{2} \grb^{2}+2\grm_{0}a^{2}(t)\geq{0}\Rightarrow a^{2}(t)\geq \frac{\grk m_{2}\grb^{2}}{2\grm_{0}}.
\end{align}
This minimum value depends on the constant value of the magnetic field $\grb$, the parameter related to the geometry $m_{2}$, and the two constants of nature involved, $\grk,\grm_{0}$.

When it comes to the charged fluid, we have studied three cases of equations of state, non-relativistic matter $(w=0)$, relativistic matter $(w=\frac{1}{3})$, and dark energy matter $(w=-1)$. In the table to follow, we present the energy density of the charged fluid, the right hand side of the equation \eqref{e99} and the effective curvature.

\begin{center}
\begin{tabular}{ |c|c|c|c| } 
\hline
\textbf{w} & $\grr^{(c)}$ & $H^{2}$ & $k_{(eff)}$ \\
\hline
0 & $-\frac{2}{\grk m_{2}a(t)^{2}}+\frac{\grr^{(c)}_{0}}{a(t)^{3}}$ & $\frac{\grk}{3}\left(\frac{\grr^{(d)}_{0}+\grr^{(c)}_{0}}{a(t)^{3}}+\frac{\grr^{(r)}_{0}}{a(t)^{4}}+\grr^{(\grL)}_{0}\right)$ & $0$ \\
\hline
$\frac{1}{3}$ & $-\frac{1}{\grk m_{2}a(t)^{2}}+\frac{\grr^{(c)}_{0}}{a(t)^{4}}$ & $\frac{\grk}{3}\left(\frac{\grr^{(d)}_{0}}{a(t)^{3}}+\frac{\grr^{(r)}_{0}+\grr^{(c)}_{0}}{a(t)^{4}}+\grr^{(\grL)}_{0}\right)+\frac{1}{3 m_{2} a(t)^{2}}$ & $-\frac{1}{3 m_{2}}<0$\\
\hline
$-1$ & $\frac{1}{\grk m_{2}a(t)^{2}}+\grr^{(c)}_{0}$ & $\frac{\grk}{3}\left(\frac{\grr^{(d)}_{0}}{a(t)^{3}}+\frac{\grr^{(r)}_{0}}{a(t)^{4}}+\grr^{(\grL)}_{0}+\grr^{(c)}_{0}\right)+\frac{1}{m_{2} a(t)^{2}}$ & $-\frac{1}{m_{2}}<0$\\
\hline
\end{tabular}
\captionof{table}{The current density $\grr^{(c)}$, the square of the Hubble function $H^{2}$ and the effective curvature $k_{(eff)}$ are presented for three cases of equations of state for the charged fluid: non-relativistic $w=0$, relativistic $w=\frac{1}{3}$ and dark energy $w=-1$. }
\end{center} 
The energy density of the electromagnetic field is the same for all the cases, $\grr^{(em)}=\frac{1}{\grk m_{2}a(t)^{2}}>0$.
For $a(t)$ positive $\forall t\in \mathbb{R}$, the term of $\grr^{(c)}$ involving the constant $\grr_{0}^{(c)}$ dominates at the limit $a(t)\rightarrow0$ for the first two cases. Thus, $\grr_{0}^{(c)}$ has to be positive in order for the energy density to be positive at that limit. In order to respect the weak energy condition \cite{Hawking:1973uf} through the whole evolution of the scale factor, an upper bound has to be imposed for these two cases.\\
\begin{align}
w=0\rightarrow a(t)^{2}\leq\left(\frac{\grk m_{2}\grr^{(c)}_{0}}{2}\right)^{2},\,\, w=\frac{1}{3}\rightarrow a(t)^{2}\leq\grk m_{2} \grr^{(c)}_{0}.
\end{align}
Considering the third case, the constant $\grr^{(c)}_{0}$ dominates at $a(t)\rightarrow \infty$ thus by use of the same argument as before, $\grr^{(c)}_{0}>0$. There is no upper bound in this case. Note also that the energy density of the charge fluid has a contribution to the usual matter with which it shares the same equation of state, for instance, when $w=0\Rightarrow \grr^{(c)}\sim \frac{\grr^{(c)}_{0}}{a(t)^{3}}$ and so on. This is due to the matter character of the fluid. Except from that, there is a contribution which scales as $\sim\frac{1}{a(t)^{2}}$ and is related to the term $\int{E^{i}J_{i}dt}$; the interaction of the charged fluid with the electromagnetic field. For the usual matter, $w=0$, $w=\frac{1}{3}$, this term has a negative sign corresponding to energy losses. On the other hand, when $w=-1$ the sign is positive; in some sense the fluid gains energy from the interaction with the electromagnetic field. A more detailed explanation is given in the Discussion.
 
\subsection{Type VIII or $A_{3,8}$ or SU$(1,1)$} 

The solutions found in this Type are separated into two cases. The structure of the text followed is almost identical to the previous Type.

\subsubsection{Case 1, $m_{2}>m_{1}$ }

\begin{align}
ds_{(4)}^{2}&=-dt^{2}+a(t)^{2}\Big[m_{1}dx^{2}-2 m_{1}sinhy\, dxdz+\left(m_{2}-m_{1}sin^{2}x\right)dy^{2} \nonumber\\
& -2m_{1}cosx\,sinx\,coshy\,dydz+\left[\left(m_{2}-m_{1} cos^{2}x\right)cosh^{2}y+m_{1}sinh^{2}y
\right]dz^{2}\Big],\\
\mathbf{F}&=-\frac{\sqrt{2 m_{2}m_{1}\grm_{0}a(t)^{2}-\grk (m_{2}-m_{1})A(t)^{2}}}{\sqrt{\grk m_{2}m_{1}a(t)^{2}}}\Big[sinx\,dt\wedge dy-cosx\,cosh(y)dt\wedge dz\Big]+\nonumber\\
& A(t) \Big[-cosx\,dx\wedge dy +sinx\,coshy\, dx\wedge dz-cosx\,sinhy\, dy\wedge dz\Big],\\
\mathbf{J}&=-\frac{2\sqrt{m_{2}m_{1}}\dot{a}(t)}{\sqrt{\grk}\sqrt{2 m_{2}m_{1}\grm_{0}a(t)^{2}-\grk (m_{2}-m_{1})A(t)^{2}}}\left[sinx\, dy+cosx\,coshy\,dz\right].
\end{align} 
The function $A(t)$ has to be determined from the following first order differential equation
\begin{align}
\dot{A}(t)=\frac{\sqrt{2 m_{2}m_{1}\grm_{0}a(t)^{2}-\grk (m_{2}-m_{1})A(t)^{2}}}{\sqrt{\grk m_{2}m_{1}a(t)^{2}}}.
\end{align}
Even though we are not able to find the analytical expression for $A(t)$ in terms of $a(t)$, that doesn't affect the ``absorption'' of the anisotropy neither the expression for $H^{2}$. This function appears only in the Faraday tensor and the four current, thus for every solution $a(t)$ of the resulting ${\grL CDM}$ equations, a solution of the previous equation will be given, if possible, analytically. The parameter $m_{3}$ was given in terms of $m_{1}$ and $m_{2}$ as $m_{3}=m_{2}-m_{1}$, hence the condition $m_{2}>m_{1}$. Same as before, there is a bound for the scale factor
\begin{align}
a(t)^{2}\geq \frac{\grk (m_{2}-m_{1}) A(t)^{2}}{2 m_{2} m_{1}\grm_{0}}.
\end{align}
Considering the effective ${\grL CDM}$ equations and the upper bounds for the scale factor of the first two cases for the equation of state, they have the same form as in Type $VI_{(-1)}$. The only difference being that the parameter $m_{2}$ of Type $VI_{(-1)}$ is replaced by the difference $m_{2}-m_{1}$ of the parameters $m_{2}, m_{1}$ of Type $VIII$. 

\begin{center}
\begin{tabular}{ |c|c|c|c| } 
\hline
\textbf{w} & $\grr^{(c)}$ & $H^{2}$ & $k_{(eff)}$ \\
\hline
0 & $-\frac{2}{\grk (m_{2}-m_{1})a(t)^{2}}+\frac{\grr^{(c)}_{0}}{a(t)^{3}}$ & $\frac{\grk}{3}\left(\frac{\grr^{(d)}_{0}+\grr^{(c)}_{0}}{a(t)^{3}}+\frac{\grr^{(r)}_{0}}{a(t)^{4}}+\grr^{(\grL)}_{0}\right)$ & $0$ \\
\hline
$\frac{1}{3}$ & $-\frac{1}{\grk (m_{2}-m_{1})a(t)^{2}}+\frac{\grr^{(c)}_{0}}{a(t)^{4}}$ & $\frac{\grk}{3}\left(\frac{\grr^{(d)}_{0}}{a(t)^{3}}+\frac{\grr^{(r)}_{0}+\grr^{(c)}_{0}}{a(t)^{4}}+\grr^{(\grL)}_{0}\right)+\frac{1}{3 (m_{2}-m_{1}) a(t)^{2}}$ & $-\frac{1}{3 (m_{2}-m_{1})}<0$\\
\hline
$-1$ & $\frac{1}{\grk (m_{2}-m_{1})a(t)^{2}}+\grr^{(c)}_{0}$ & $\frac{\grk}{3}\left(\frac{\grr^{(d)}_{0}}{a(t)^{3}}+\frac{\grr^{(r)}_{0}}{a(t)^{4}}+\grr^{(\grL)}_{0}+\grr^{(c)}_{0}\right)+\frac{1}{(m_{2}-m_{1}) a(t)^{2}}$ & $-\frac{1}{(m_{2}-m_{1})}<0$\\
\hline
\end{tabular}
\captionof{table}{The current density $\grr^{(c)}$, the square of the Hubble function $H^{2}$ and the effective curvature $k_{(eff)}$ are presented for three cases of equations of state for the charged fluid: non-relativistic $w=0$, relativistic $w=\frac{1}{3}$ and dark energy $w=-1$. }
\end{center}
\begin{align}
w=0\rightarrow a(t)^{2}\leq\left(\frac{\grk (m_{2}-m_{1})\grr^{(c)}_{0}}{2}\right)^{2},\,\, w=\frac{1}{3}\rightarrow a(t)^{2}\leq\grk (m_{2}-m_{1}) \grr^{(c)}_{0}.
\end{align}
The energy density of the electromagnetic field is $\grr^{(em)}=\frac{1}{\grk(m_{2}-m_{1})a(t)^{2}}>0$.

\subsubsection{Case 2, $m_{3}>m_{1}$ }

\begin{align}
ds_{(4)}^{2}&=-dt^{2}+a(t)^{2}\Big[m_{1}dx^{2}-2 m_{1}sinhy\,dxdz+\left(m_{3}-m_{1}cos^{2}x\right)dy^{2} \nonumber\\
& +2m_{1}cosx\,sinx\,coshy\,dydz+\left[\left(m_{3}-m_{1} sin^{2}x\right)cosh^{2}y+m_{1}sinh^{2}y
\right]dz^{2}\Big],\\
\mathbf{F}&=-\frac{\sqrt{2 m_{3}m_{1}\grm_{0}a(t)^{2}-\grk (m_{3}-m_{1})A(t)^{2}}}{\sqrt{\grk m_{3}m_{1}a(t)^{2}}}\Big[cosx\,dt\wedge dy-sinx\,coshy\,dt\wedge dz\Big]+\nonumber\\
& A(t) \Big[sinx\,dx\wedge dy +cosx\,coshy\, dx\wedge dz+sinx\, sinhy\, dy\wedge dz\Big],\\
\mathbf{J}&=-\frac{2\sqrt{m_{3}m_{1}}\dot{a}(t)}{\sqrt{\grk}\sqrt{2 m_{3}m_{1}\grm_{0}a(t)^{2}-\grk (m_{3}-m_{1})A(t)^{2}}}\left[cosx\, dy-sinx\,coshy\,dz\right].
\end{align} 
The function $A(t)$ is determined from the same equation as before with $m_{2}$ replaced by $m_{3}$. The parameter $m_{2}$ was given in terms of $m_{1}$ and $m_{3}$ as $m_{2}=m_{3}-m_{1}$, hence the condition $m_{3}>m_{1}$. The maximum values for the scale factor and the effective equations are the same with the ones of the previous case, where $m_{2}$ has to be replaced by $m_{3}$. This metric as well as the previous one, admits only the original three Killing fields of the Bianchi Type VIII.
 
\subsection{Type IX or $A_{3,9}$ or SU$(2)$} 

Finally, we present the solutions found for Type IX. There are three families in this Type.

\subsubsection{Case 1, $m_{2}>m_{3}$, $m_{3}=m_{2}-m_{4}$, $m_{4}>0$ }

\begin{align}
ds_{(4)}^{2}&=-dt^{2}+a(t)^{2}\Big[m_{2}dx^{2}+2 m_{2}siny\,dxdz+\left(m_{2}-m_{4}sin^{2}x\right)dy^{2} \nonumber\\
& -2m_{4}cosx\,sinx\,cosy\,dydz+\left[m_{2}-m_{4} cos^{2}x\,cos^{2}y
\right]dz^{2}\Big],\\
\mathbf{F}&=-\frac{\sqrt{m_{2}-m_{4}}\sqrt{m_{4}\grm_{0}a(t)^{2}-\grk A(t)^{2}}}{\sqrt{\grk}m_{2}a(t)}\Big[sinx\,dt\wedge dy+cosx\,cosy\,dt\wedge dz\Big]+\nonumber\\
& A(t) \Big[-cosx\,dx\wedge dy +sinx\,cosy\, dx\wedge dz+cosx\, siny\, dy\wedge dz\Big],\\
\mathbf{J}&=-\frac{m_{4}\sqrt{m_{2}-m_{4}}\dot{a}(t)}{\sqrt{\grk}m_{2}\sqrt{m_{4}\grm_{0}a(t)^{2}-\grk A(t)^{2}}}\left[sinx\, dy+cosx\,cosy\,dz\right].
\end{align} 
The function $A(t)$ has to be determined from the following first order equation differential equation $\dot{A}(t)=\frac{\sqrt{m_{2}-m_{4}}\sqrt{m_{4}\grm_{0}a(t)^{2}-\grk A(t)^{2}}}{\sqrt{\grk}m_{2}a(t)}$. The parameter $m_{4}$ was introduced for the simplification of the expressions, while $m_{1}$ was equal to $m_{2}$. Same as before, there is a bound for the scale factor
\begin{align}
a(t)^{2}\geq \frac{\grk A(t)^{2}}{m_{4}\grm_{0}}.
\end{align}
In the Table below, we present the effective equations 

\begin{center}
\begin{tabular}{ |c|c|c|c| } 
\hline
\textbf{w} & $\grr^{(c)}$ & $H^{2}$ & $k_{(eff)}$ \\
\hline
$0$ & $-\frac{m_{4}}{\grk m_{2}^{2}a(t)^{2}}+\frac{\grr^{(c)}_{0}}{a(t)^{3}}$ & $\frac{\grk}{3}\left(\frac{\grr^{(d)}_{0}+\grr^{(c)}_{0}}{a(t)^{3}}+\frac{\grr^{(r)}_{0}}{a(t)^{4}}+\grr^{(\grL)}_{0}\right)-\frac{m_{2}+m_{4}}{4m_{2}^{2}a(t)^{2}}$ & $\frac{m_{2}+m_{4}}{4m_{2}^{2}}>0$ \\
\hline
$\frac{1}{3}$ & $-\frac{m_{4}}{2\grk m_{2}^{2}a(t)^{2}}+\frac{\grr^{(c)}_{0}}{a(t)^{4}}$ & $\frac{\grk}{3}\left(\frac{\grr^{(d)}_{0}}{a(t)^{3}}+\frac{\grr^{(r)}_{0}+\grr^{(c)}_{0}}{a(t)^{4}}+\grr^{(\grL)}_{0}\right)-\frac{3 m_{2}+m_{4}}{12 m_{2}^{2}a(t)^{2}}$ & $\frac{3 m_{2}+m_{4}}{12 m_{2}^{2}}>0$ \\
\hline
$-1$ & $\frac{m_{4}}{2\grk m_{2}^{2}a(t)^{2}}+\grr^{(c)}_{0}$ & $\frac{\grk}{3}\left(\frac{\grr^{(d)}_{0}}{a(t)^{3}}+\frac{\grr^{(r)}_{0}}{a(t)^{4}}+\grr^{(\grL)}_{0}+\grr^{(c)}_{0}\right)-\frac{m_{2}-m_{4}}{4m_{2}^{2}a(t)^{2}}$ & $\frac{m_{2}-m_{4}}{4m_{2}^{2}}>0$\\
\hline
\end{tabular}
\captionof{table}{The current density $\grr^{(c)}$, the square of the Hubble function $H^{2}$ and the effective curvature $k_{(eff)}$ are presented for the three cases of equations of state for the charged fluid: non-relativistic $w=0$, relativistic $w=\frac{1}{3}$ and dark energy $w=-1$. }
\end{center}
The energy density of the electromagnetic field is given by $\grr^{(em)}=\frac{m_{4}}{2\grk m_{2}^{2}a(t)^{2}}>0$.
Note that this metric admits only the three original Killing fields. Another important feature is the non-zero effective curvature when $w=0$ and the fact that the effective curvature is always positive.  Nevertheless, the scale factor has to be bounded for the first two cases 
\begin{align}
w=0\rightarrow a(t)^{2}\leq\left(\frac{\grk m_{2}^{2}\grr^{(c)}_{0}}{m_{4}}\right)^{2},\,\,w=\frac{1}{3}\rightarrow a(t)^{2}\leq\frac{2\grk m_{2}^{2}\grr^{(c)}_{0}}{m_{4}}.
\end{align}

\subsubsection{Case 2, $m_{3}>m_{2}$, $m_{2}=m_{3}-m_{4}$, $m_{4}>0$ }

\begin{align}
ds_{(4)}^{2}&=-dt^{2}+a(t)^{2}\Big[m_{3}dx^{2}+2 m_{3}siny\,dxdz+\left(m_{3}-m_{4}cos^{2}x\right)dy^{2}+ \nonumber\\
& 2m_{4}cosx\,sinx\,cosy\,dydz+\left[m_{3}-m_{4} sin^{2}x\,cos^{2}y
\right]dz^{2}\Big],\\
\mathbf{F}&=-\frac{\sqrt{m_{3}-m_{4}}\sqrt{m_{4}\grm_{0}a(t)^{2}-\grk A(t)^{2}}}{\sqrt{\grk}m_{3}a(t)}\Big[cosx\,dt\wedge dy-sinx\,cosy\,dt\wedge dz\Big]+\nonumber\\
& A(t) \Big[sinx\,dx\wedge dy +cosx\,cosy\, dx\wedge dz-sinx\, siny\, dy\wedge dz\Big],\\
\mathbf{J}&=-\frac{m_{4}\sqrt{m_{3}-m_{4}}\dot{a}(t)}{\sqrt{\grk}m_{3}\sqrt{m_{4}\grm_{0}a(t)^{2}-\grk A(t)^{2}}}\left[cosx\, dy-sinx\,cosy\,dz\right].
\end{align} 
The function $A(t)$, the boundaries and the effective equations are the same as previously with the only difference being the replacement of $m_{2}$ from $m_{3}$. The original parameter $m_{1}$ was equal to $m_{3}$. This metric admits the same Killing fields as the previous one.
 
\subsubsection{Case 3, $m_{3}>m_{1}$, $m_{3}=m_{1}+m_{4}$, $m_{4}>0$ }

\begin{align}
ds_{(4)}^{2}=&-dt^{2}+a(t)^{2}\Big[m_{1}dx^{2}+2 m_{1}siny\,dxdz+(m_{1}+m_{4})dy^{2}
+\left[m_{1}+m_{4}cos^{2}y\right]dz^{2}\Big],
\end{align}
\begin{align}
\mathbf{F}=&-\frac{\sqrt{m_{1}}\sqrt{m_{4}\grm_{0}a(t)^{2}-\grk A(t)^{2}}}{\sqrt{\grk}(m_{1}+m_{4})a(t)}\Big[dt\wedge dx+siny\,dt\wedge dz\Big]-A(t)cosy\,dy\wedge dz,\\
\mathbf{J}=&-\frac{\sqrt{m_{1}}m_{4}\dot{a}(t)}{\sqrt{\grk}(m_{1}+m_{4})\sqrt{m_{4}\grm_{0}a(t)^{2}-\grk A(t)^{2}}}\left[dx+siny\,dz\right].
\end{align} 
The function $A(t)$ has to be determined from the following equation $\dot{A}(t)=\frac{\sqrt{m_{1}}\sqrt{m_{4}\grm_{0}a(t)^{2}-\grk A(t)^{2}}}{\sqrt{\grk}(m_{1}+m_{4})a(t)}$. The parameter $m_{2}$ was equal to $m_{3}$. The same bound for the scale factor holds as before.
In the Table below, the effective equations are presented.

\begin{center}
\begin{tabular}{ |c|c|c|c| } 
\hline
\textbf{w} & $\grr^{(c)}$ & $H^{2}$ & $k_{(eff)}$ \\
\hline
$0$ & $-\frac{m_{4}}{\grk (m_{1}+m_{4})^{2}a(t)^{2}}+\frac{\grr^{(c)}_{0}}{a(t)^{3}}$ & $\frac{\grk}{3}\left(\frac{\grr^{(d)}_{0}+\grr^{(c)}_{0}}{a(t)^{3}}+\frac{\grr^{(r)}_{0}}{a(t)^{4}}+\grr^{(\grL)}_{0}\right)-\frac{m_{1}+2m_{4}}{4(m_{1}+m_{4})^{2}a(t)^{2}}$ & $\frac{m_{1}+2m_{4}}{4(m_{1}+m_{4})^{2}}>0$ \\
\hline
$\frac{1}{3}$ & $-\frac{m_{4}}{2\grk (m_{1}+m_{4})^{2}a(t)^{2}}+\frac{\grr^{(c)}_{0}}{a(t)^{4}}$ & $\frac{\grk}{3}\left(\frac{\grr^{(d)}_{0}}{a(t)^{3}}+\frac{\grr^{(r)}_{0}+\grr^{(c)}_{0}}{a(t)^{4}}+\grr^{(\grL)}_{0}\right)-\frac{3 m_{1}+4 m_{4}}{12 (m_{1}+m_{4})^{2}a(t)^{2}}$ & $\frac{3 m_{1}+4 m_{4}}{12 (m_{1}+m_{4})^{2}}>0$ \\
\hline
$-1$ & $\frac{m_{4}}{2\grk (m_{1}+m_{4})^{2}a(t)^{2}}+\grr^{(c)}_{0}$ & $\frac{\grk}{3}\left(\frac{\grr^{(d)}_{0}}{a(t)^{3}}+\frac{\grr^{(r)}_{0}}{a(t)^{4}}+\grr^{(\grL)}_{0}+\grr^{(c)}_{0}\right)-\frac{m_{1}}{4(m_{1}+m_{4})^{2}a(t)^{2}}$ & $\frac{m_{1}}{4(m_{1}+m_{4})^{2}}>0$\\
\hline
\end{tabular}
\captionof{table}{The current density $\grr^{(c)}$, the square of the Hubble function $H^{2}$ and the effective curvature $k_{(eff)}$ are presented for three cases of equations of state for the charged fluid: non-relativistic $w=0$, relativistic $w=\frac{1}{3}$ and dark energy $w=-1$. }
\end{center}
Finally, the electromagnetic field has energy density $\grr^{(em)}=\frac{m_{4}}{2\grk (m_{1}+m_{4})^{2}a(t)^{2}}>0$.
Note also that this metric admits one more Killing field, $\grz=\partial_{x}$, which implies that belongs to the family of Locally Rotationally Symmetric (LRS) sub-class of the Type treated. The maximum values for the first two cases read
\begin{align}
w=0\rightarrow a(t)^{2}\leq\left(\frac{\grk (m_{1}+m_{4})^{2}\grr^{(c)}_{0}}{m_{4}}\right)^{2},\,\,w=\frac{1}{3}\rightarrow a(t)^{2}\leq\frac{2\grk (m_{1}+m_{4})^{2}\grr^{(c)}_{0}}{m_{4}}.
\end{align}
 
\section{Discussion} 
\label{sec6}
 
In the present work, we have investigated whether, under specific assumptions, dynamically effective equations equivalent to those of $\grL CDM$, could result from Bianchi spacetimes. The primary assumption is that the spatial metric is a constant matrix $m_{\grm\grn}$ (encompassing the ``frozen'' anisotropy) multiplied by one time-dependent function $a(t)^{2}$ (taking the role of the scale factor in the FLRW models). This implies the existence of a conformal Killing vector field which is proportional to the comoving velocity vector field. From this follows that the spacetime is parallax free and the temperature (assuming black body spectrum) of the comoving radiation fluid does not depend on the direction of observation \cite{1988JMP....29.2064H}, \cite{1992GReGr..24..121O}. The next step was to search for fields whose energy momentum tensor could be physical and capable of ``absorbing'' this anisotropy; thus, effectively only one constraint and the corresponding dynamical equation for $a(t)$ results. We found that, in order for this to be the case, the flux of the total matter content should be zero, while the traceless anisotropic pressure tensor should be a function of the spatial Ricci tensor and the corresponding Ricci scalar. 

An important tool in our analysis is the group of constant Automorphisms,  used to transform the matrix $m_{\grm\grn}$ in an irreducible form, i.e. containing only the essential constants characterizing the space in question. We present for each Type the Automorphism matrix ${\grL^{\grm}}_{\grn}$ and the corresponding matrix $m_{\grm\grn}$: Bianchi Type $I$ has no essential constant left, Types $II$, $V$ have one, the Types $VIII$, $IX$ have three, while the rest of them ($III$, $IV$, $VI_{(h)}$, $VII_{(h)}$) have two. The maximum number of essential constants belongs to Types $VIII$, $IX$ so, in some sense, these are the most general geometries in the set. The final form of the metric is the simplest possible without loss of generality. We believe that this could prove to be helpful when the solutions are compared with observations, since the remaining constants are essential and thus are related to quantities which are physically interesting. For instance, in the case of Type $VI_{(-1)}$ the only remaining constant $m_{2}$ is related to the effective curvature $k_{eff}$. On the contrary, if we had not use the group of Automorphisms, there would be non-essential constants present and would be difficult to isolate the constants with the physical meaning.

By assuming the existence of one free scalar field we reproduce the only already known solution. Our form of the solution has only one essential constant and we claim that this is the most general one that can be found. This parameter appears in the effective curvature $k_{(eff)}=-\frac{1}{2m_{2}}$ and since $m_{2}$ is strictly positive (in order for the spacetime metric to have Lorentz signature) $k_{(eff)}<0$.
 
We next try to ``absorb'' the anisotropy via an electromagnetic field. This can be seen not to work for a free electromagnetic field; thus we also assume the existence of a charged fluid interacting with it. A secondary assumption is that the charged fluid has the form of a perfect fluid with equation of state $P^{(c)}=w \grr^{(c)}$. Hence, the total flux and
traceless anisotropic pressure tensor has to be equated with those of the electromagnetic field. Three cases are investigated, non-relativistic $(w=0)$, relativistic $(w=\frac{1}{3})$ and dark energy-like $(w=-1)$ fluid. The only solutions found belong to the Types $VI_{(-1)}$, $VIII$, $IX$. 

In relation to the case of Type $VI_{(-1)}$: from the two original geometric parameters only one is left. In order for the solution to be valid, the scale factor $a(t)^{2}$ should be bounded from below. This bound depends on the remaining geometric parameter, the constant magnetic field of the solution and the two constants of nature involved. This indicates a universe with no initial singularity (big bang). The effective curvature is zero when $w=0$; it is thus dynamically equivalent to a flat FLRW. The other two cases are equivalent to an open one, $k_{(eff)}<0$. In both cases the effective curvature contains the geometric parameter. Furthermore, in the cases $w=0$, $w=\frac{1}{3}$, in order for the energy density of the charge fluid to be positive, there has to be an upper bound for the scale factor, different in each case. This bound depends on the $\grk$ and some constant related to the energy density of the charged fluid. The solution admits only the three Killing fields related to the slice's homogeneity.

For the Type $VIII$, two cases came up: from the three original parameters only two are left in each case. A first order differential equation for some function $A(t)$ is left to be solved in order for the explicit form of the Faraday tensor and the four current to be given. This however does not affect the whole analysis neither the ``absorption'' of the anisotropy. This function is linked to the magnetic field. There is again a lower bound which depends on the two geometric parameters, the function $A(t)$ and the two constants $\grk,\,\grm_{0}$. Once more, for $w=0$ an equivalent flat FLRW came up, while the other two cases correspond to open FLRW. In the effective curvature of the open spaces, only the difference of the two geometric parameters appears. An upper boundary for the scale factor exists as well, when $w=0$ and $w=\frac{1}{3}$ (again for the weak energy condition to be satisfied). The Killing fields of the resulting metric are the three of the Bianchi Type.
  
Finally, in Type IX there are three cases: in each case two geometric parameters are left. As in Type VIII, there is a function to be determined. There is a lower bound as in the two previous Types. For each case and type of fluid, the effective curvature is positive, thus corresponds to closed FLRW. In accordance with the previous results there are upper bounds for the scale factor when we consider non-relativistic and relativistic charged fluid. In the third case, an additional Killing field exists with the resulting metric belonging to the Locally Rotationally Symmetric (LRS) sub-class of the treated model.

We have found all the classes of (effective) FLRW spacetimes, closed, open and flat. However, this came at a cost, namely the existence of upper bounds for the scale factor. Should the corresponding solutions be considered as non physical? If the charge energy density was strictly negative, then the answer would certainly be yes. In the cases at hand the dependence of the $\grr^{(c)}$ on the scale factor is such that, although initially positive, when $a(t)$ crosses the given bounds the energy density becomes negative. The explanation for this occurrence can be linked to the balance between the matter character of the fluid which scales as $\frac{1}{a(t)^{3}}$ or $\frac{1}{a(t)^{4}}$ (depending on whether we consider non-relativistic or relativistic charged fluid), and the charge character which scales as $-\frac{1}{a(t)^{2}}$ and is related to the term $\int{E^{i}J_{i}\,dt}$. The problem of negative energy density is localized in the combination of the negative sign and the power of the scaling of the term $\int{E^{i}J_{i}\,dt}$. The minus sign can be understood as energy losses due to the interaction with the electromagnetic field. To make this clear, let us consider the case of $w=0$ for Type $VI_{(-1)}$ and at the same time assume that the scale factor lies within the acceptable bounds for a specific value of time $a(t)=a(t_{0})$: for the uncharged fluid the energy density would be
\begin{align}
\grr^{(d)}=\frac{\grr_{0}}{a(t_{0})^{3}},
\end{align}
while for the charged fluid with the same equation of state would be
\begin{align}
\grr^{(c)}=\frac{\grr_{0}}{a(t_{0})^{3}}-\frac{2}{\grk m_{2} a(t_{0})^{2}},
\end{align}
where we have considered the case in which the two fluids have equal original energy densities $\grr_{0}$. It is rather obvious that $\grr^{(c)}<\grr^{(d)}$ thus some portion of energy has been spend on the interaction. The power of the scaling for the term $\int{E^{i}J_{i}\,dt}$ is related to the ``absorption'' of the anisotropy and that is why is the same in all the types of equation of state. To conclude this paragraph, for the weak energy condition to be fulfilled, these solutions should be considered physical only for specific time (and thus scale factor) interval.
 
A charged fluid which shares the same equation of state with a dark energy fluid has been considered as well. In contrast to the previous cases, there is no negative energy density problem, and that is due to the positive sign of the term $\frac{1}{a(t)^{2}}$. Thus, the charged fluid with this equation of state seems to gain energy from the interaction with the electromagnetic field. The shortcoming of this case is that there is no solution with exactly zero effective curvature, although there are parameters left which may be fine tuned in order for the effective curvature to be considered as almost zero. 

Let us also point out another fact: the Type $VI_{(-1)}$ is considered, where the first Friedmann equation and the charged energy density for the case of dark energy-like fluid reads 
\begin{align}
&H^{2}=\frac{\grk}{3}\left(\frac{\grr^{(d)}_{0}}{a(t)^{3}}+\frac{\grr^{(r)}_{0}}{a(t)^{4}}+\grr_{0}^{(\grL)}+\grr_{0}^{(c)}\right)+\frac{1}{m_{2}a(t)^{2}},\\
&\grr^{(c)}=\frac{1}{\grk m_{2} a(t)^{2}}+\grr^{(c)}_{0}.
\end{align}
The ability exists to ignore the uncharged cosmological constant fluid and identify as $\grr_{0}^{(\grL)}$ the $\grr_{0}^{(c)}$; the equations would be once more dynamically equivalent to those of $\grL CDM$ with open FRLW underlying geometry.
\begin{align}
H^{2}&=\frac{\grk}{3}\left(\frac{\grr^{(d)}_{0}}{a(t)^{3}}+\frac{\grr^{(r)}_{0}}{a(t)^{4}}+\grr_{0}^{(\grL)}\right)+\frac{1}{m_{2}a(t)^{2}},\\
\grr^{(c)}&=\frac{1}{\grk m_{2} a(t)^{2}}+\grr^{(\grL)}_{0}.
\end{align}
Thus, for $a(t)\rightarrow 0$ the charged energy density scales as $\frac{1}{\grk m_{2} a(t)^{2}}$, in other words like the effective curvature, while for $a(t)\rightarrow \infty$ behaves like the energy density of the cosmological constant fluid. Therefore, the possibility for the observed dark energy fluid to be charged should be considered. 

Let us now look at the form of the norm of charged current density $J^{i}$, in the case of Type $VI_{(-1)}$ where the explicit form is given.
\begin{align}
|J|=\sqrt{\frac{4}{\grk m_{2}\left(-\grk m_{2} \grb^{2}+2 \grm_{0} a(t)^{2}\right)}}H(t).
\end{align}
Away from the lower bound 
\begin{align}
|J|\sim \frac{1}{a(t)}H(t).
\end{align}
If for instance, we study the radiation dominated era, $H^{2}\sim \frac{1}{a(t)^{4}}\Rightarrow|J|\sim \frac{1}{a(t)^{3}}$ thus it scales like non-relativistic matter fluid. Hence, imprints of this current density may be found in the CMB, but we need to be more thorough-full with this thought in some future work.

At the mathematical level we consider that this paper points to a negative answer to the question raised in the introduction ``Does the observational data of the CMB uniquely fix the spacetime metric?'', with every restriction that we discussed above.

Finally, let us point out some further  ideas for future work.
\begin{enumerate}
\item We can search for coordinate transformations that will transform the metrics into a more convenient form in order to estimate all the parameters of the models found. The next step would be to compare the results with the well known $\grL CDM$ where the underlying geometry is the flat FLRW  and find out what is the level of agreement, if any, with the observational data.
\item Perturbations upon the solutions found would also be of interest.
\item It would be interesting to search for other spacetimes where this process of positive to negative energy densities takes place. 
\end{enumerate} 

\begin{appendices}

\section{Automorphisms}

\subsection{Time dependent Automorphisms}

In \cite{2002CMaPh.226..377C}, \cite{doi:10.1063/1.1386637} a group of coordinate transformations was derived that preserves the line element's manifest homogeneity and, as a side effect, generate symmetries of the corresponding Einstein's field equations. We briefly recall the idea:

Let us start with the most general form of a line element which admits a three dimensional isometry group G acting simply transitively on the hypersurfaces $t=\text{constant}$; the corresponding Killing fields to be applied on the metric are the trivial prolongations $\grj_{\grm}=0\partial_{t}+\grj_{\grm}^{i}\partial_{i}$. As we have argued in section (2.1), there exist and invariant basis of one-forms $\grs^{\gra}$, such that (in a 3+1 decomposition) the metric assumes the form
\begin{align}
ds_{(4)}^{2}=\left[-N(t)^{2}+N_{\grm}(t)N^{\grm}(t)\right]dt^{2}+2N_{\grm}(t)\grs^{\grm}_{i}(x)dt\,dx^{i}+\grg_{\grm\grn}(t)\grs^{\grm}_{i}(x)\grs^{\grn}_{j}(x)dx^{i}\,dx^{j},\label{lefth0}
\end{align}
where $(\grm,\grn,..)$ are triad indices while $(i,j,...)$ coordinate basis indices. All of them run from $1$ to $3$. One can still search for Gaussian normal coordinates, the existence of which was used in section 2.1, in order to arrive at \eqref{eqs15}; then the question would arise as to whether the transformation needed would preserve the manifest spatial homogeneity of \eqref{lefth0} or not. As we shall see, the answer is positive.

Now, let us perform the following coordinate transformations
\begin{align}
t\mapsto\tilde{t}=t, \hspace{2.0cm}\\
x^{i}\mapsto{\tilde{x}}^{i}=h^{i}(t,x^{l}), \hspace{0.2cm}x^{i}=f^{i}(\tilde{t},{\tilde{x}}^{l}),
\end{align}
which, after some mathematical manipulations, results in the new form of the initial line element
\begin{align}
ds_{(4)}^{2}&=\Bigg\{\left[-N(\tilde{t})^{2}+N_{\grm}(\tilde{t})N^{\grm}(\tilde{t})\right]+2N_{\grm}(\tilde{t})\grs^{\grm}_{i}(f)\frac{\partial f^{i}}{\partial \tilde{t}}+\grg_{\grm\grn}(\tilde{t})\grs^{\grm}_{i}(f)\frac{\partial f^{i}}{\partial \tilde{t}}\grs^{\grn}_{j}(f)\frac{\partial f^{j}}{\partial \tilde{t}}\Bigg\}d\tilde{t}^{2}+\nonumber\\
&2\left[N_{\grm}(\tilde{t})\grs^{\grm}_{i}(f)\frac{\partial f^{i}}{\partial \tilde{x}^{m}}+\grg_{\grm\grn}(\tilde{t})\grs^{\grm}_{i}(f)\frac{\partial f^{i}}{\partial \tilde{t}}\grs^{\grn}_{j}(f)\frac{\partial f^{j}}{\partial \tilde{x}^{m}}\right]d\tilde{t}d\tilde{x}^{m}+\grg_{\grm\grn}(\tilde{t})\grs^{\grm}_{i}(f)\frac{\partial f^{i}}{\partial \tilde{x}^{m}}\grs^{\grn}_{j}(f)\frac{\partial f^{i}}{\partial \tilde{x}^{n}}d\tilde{x}^{m}d\tilde{x}^{n}, \label{lefth}
\end{align}
which by introducing the abbreviations
\begin{align}
&\grs^{\grm}_{i}(f)\frac{\partial f^{i}}{\partial \tilde{t}}=P^{\grm}(\tilde{t},\tilde{x}),\label{sigma}\\
&\grs^{\grm}_{i}(f)\frac{\partial f^{i}}{\partial \tilde{x}^{m}}={\grL^{\grm}}_{m}(\tilde{t},\tilde{x}),\label{migma}
\end{align}
becomes
\begin{align}
ds_{(4)}^{2}&=\Bigg\{\left[-N(\tilde{t})^{2}+N_{\grm}(\tilde{t})N^{\grm}(\tilde{t})\right]+2N_{\grm}(\tilde{t})P^{\grm}(\tilde{t},\tilde{x})+\grg_{\grm\grn}(\tilde{t})P^{\grm}(\tilde{t},\tilde{x})P^{\grn}(\tilde{t},\tilde{x})\Bigg\}d\tilde{t}^{2}+\nonumber\\
&2\left[N_{\grm}(\tilde{t}){\grL^{\grm}}_{m}(\tilde{t},\tilde{x})+\grg_{\grm\grn}(\tilde{t})P^{\grm}(\tilde{t},\tilde{x}){\grL^{\grn}}_{m}(\tilde{t},\tilde{x})\right]d\tilde{t}d\tilde{x}^{m}+\grg_{\grm\grn}(\tilde{t}){\grL^{\grm}}_{m}(\tilde{t},\tilde{x}){\grL^{\grn}}_{n}(\tilde{t},\tilde{x})d\tilde{x}^{m}d\tilde{x}^{n}.\label{lefth1}
\end{align}
Of course, \eqref{lefth1} admits the same symmetry group as \eqref{lefth0}. At this point, we introduce the main requirement that we seek transformations which preserve the manifest spatial homogeneity. This means that the matrix ${\grL^{\grm}}_{\grn}$ and the triplet $P^{\grm}$ should satisfy the equations
\begin{align}
&{\grL^{\grm}}_{m}(\tilde{t},\tilde{x})={\grL^{\grm}}_{\grn}(\tilde{t})\grs^{\grn}_{m}(\tilde{x}),\label{lm1}\\
&P^{\grm}(\tilde{t},\tilde{x})=P^{\grm}(\tilde{t}).\label{lm2}
\end{align}
The restriction \eqref{lm1} simply states that we introduce the old basis of one-forms evaluated in the new coordinate system. The equations \eqref{sigma}, \eqref{migma} become
\begin{align}
&\grs^{\grm}_{i}(f)\frac{\partial f^{i}}{\partial \tilde{t}}=P^{\grm}(\tilde{t}),\label{sigma1}\\
&\grs^{\grm}_{i}(f)\frac{\partial f^{i}}{\partial \tilde{x}^{m}}={\grL^{\grm}}_{\grn}(\tilde{t})\grs^{\grn}_{m}(\tilde{x}),\label{migma1}
\end{align}
while the line element \eqref{lefth1} after some mathematical manipulations acquires the form,
\begin{align}
ds_{(4)}^{2}=\left[-\tilde{N}(\tilde{t})+\tilde{N}_{\grm}(\tilde{t})\tilde{N}^{\grm}(\tilde{t})\right]d\tilde{t}^{2}+2\tilde{N}_{\grm}(\tilde{t})\grs^{\grm}_{m}(\tilde{x})d\tilde{t}d\tilde{x}^{m}+\tilde{\grg}_{\grm\grn}(\tilde{t})\grs^{\grm}_{m}(\tilde{x})\grs^{\grn}_{n}(\tilde{x})d\tilde{x}^{m}d\tilde{x}^{n}
\end{align}
The following abbreviations were used
\begin{align}
\tilde{\grg}_{\gra\grb}(\tilde{t})&=\grg_{\grm\grn}(\tilde{t}){\grL^{\grm}}_{\gra}(\tilde{t}){\grL^{\grn}}_{\grb}(\tilde{t}),\label{mn1}\\
\tilde{N}_{\gra}(\tilde{t})&=\left(N_{\grm}(\tilde{t})+\grg_{\grm\grn}(\tilde{t})P^{\grn}(\tilde{t})\right){\grL^{\grm}}_{\gra}(\tilde{t}),\label{mn2}\\
\tilde{N}\left(\tilde{t}\right)&=N\left(\tilde{t}\right).\label{mn3}
\end{align}
Since the system of \eqref{sigma1}, \eqref{migma1} comprises twelve, first order, highly non-linear partial differential equations in terms of $f^{i}$, it is not at all clear whether solutions will exist or not for some ${\grL^{\grm}}_{\gra}(\tilde{t})$, $P^{\grn}(\tilde{t})$. Let us rewrite \eqref{sigma1}, \eqref{migma1}, multiplying by $\grs^{i}_{\grm}(f)$(the inverse of $\grs^{\grm}_{i}(f)$)
\begin{align}
&\frac{\partial f^{i}}{\partial \tilde{t}}=\grs^{i}_{\grm}(f)P^{\grm}(\tilde{t}),\label{fro1}\\
&\frac{\partial f^{i}}{\partial \tilde{x}^{m}}=\grs^{i}_{\grm}(f){\grL^{\grm}}_{\grn}(\tilde{t})\grs^{\grn}_{m}(\tilde{x}).\label{fro2}
\end{align}
The existence of local solutions to the equations \eqref{fro1}, \eqref{fro2} is guaranteed by the Frobenious theorem if the following necessary and sufficient conditions hold:
\begin{align}
&\frac{\partial^{2}f^{i}}{\partial \tilde{x}^{m}\partial \tilde{t}}-\frac{\partial^{2}f^{i}}{\partial \tilde{t}\partial \tilde{x}^{m}}=0,\label{fro11}\\
&\frac{\partial^{2}f^{i}}{\partial \tilde{x}^{m}\partial \tilde{x}^{n}}-\frac{\partial^{2}f^{i}}{\partial \tilde{x}^{n}\partial \tilde{x}^{m}}=0,\label{fro22}
\end{align}
which after manipulations and use of previous relations result in the following restrictions on ${\grL^{\grm}}_{\grn}$, $P^{\grm}$
\begin{align}
&{\Lambda^{\alpha}}_{\mu}\left(\tilde{t}\right)C^{\mu}_{\beta\nu}=C^{\alpha}_{\mu\sigma}{\Lambda^{\mu}}_{\beta}\left(\tilde{t}\right){\Lambda^{\sigma}}_{\nu}\left(\tilde{t}\right),\label{e34}\\
&{{\dot{\Lambda}}^{\alpha}}_{\!\!\!\quad\beta}\left(\tilde{t}\right)={\Lambda^{\mu}}_{\beta}\left(\tilde{t}\right)C^{\alpha}_{\mu\nu}P^{\nu}\left(\tilde{t}\right),\label{e35}
\end{align}
where $C^{\grm}_{\grn\grs}$ the structure constants of the isometry group. The solutions to the above integrability conditions form the group of time dependent Automporphisms. In every Bianchi Type, they contain three arbitrary functions of time and some constants of the ``rigid'' Automorphisms group(remaining gauge symmetry). This can be seen by noting that $P^{\grn}(\tilde{t})=0$, ${\grL^{\gra}}_{\grb}(\tilde{t})={\grL^{\gra}}_{\grb}$ always consist a solution. The three arbitrary functions are distributed both in $P^{\grn}(\tilde{t})$ and ${\grL^{\gra}}_{\grb}(\tilde{t})$, thus one can always use them to arrive at a zero shift $\tilde{N}_{\grm}(\tilde{t})$ through \eqref{mn2}. Therefore, we manage to find Gaussian normal coordinates and at the same time maintain the manifest spatial homogeneity.

\subsection{${\grL^{\grm}}_{\grn}$ for Bianchi Type III}

Let us start with some general constant matrix ${\grL^{\grm}}_{\grn}$
\begin{align}
{\grL^{\grm}}_{\grn}=\begin{pmatrix}
a_{1} & a_{2} & a_{3}\\
a_{4} & a_{5} & a_{6}\\
a_{7} & a_{8} & a_{9}
\end{pmatrix}.
\end{align}
By use of the structure constants for the Bianchi Type III, $C^{1}_{13}=-C^{1}_{31}=-1$, the set of independent equations \eqref{es34} becomes
\begin{align}
&a_{4}=0,\label{e123}\\
&a_{7}=0,\label{e124}\\
&a_{2}a_{7}-a_{1}a_{8}=0,\label{e125}\\
&a_{1}+a_{3}a_{7}-a_{1}a_{9}=0,\label{e126}\\
&a_{3}a_{8}-a_{2}a_{9}=0.\label{e127}
\end{align}
By using $\eqref{e123}$, $\eqref{e124}$ in the rest of the equations we acquire
\begin{align}
&a_{1}a_{8}=0,\label{e128}\\
&a_{1}-a_{1}a_{9}=0,\label{e129}\\
&a_{3}a_{8}-a_{2}a_{9}=0.\label{e130}
\end{align}
There are four sets of possible solutions to the equations $\eqref{e128}$, $\eqref{e129}$, $\eqref{e130}$:
\begin{align}
&s_{1}=\left\{a_{1}=0, a_{3}=\frac{a_{2}a_{9}}{a_{8}}\right\},\\
&s_{2}=\left\{a_{1}=0, a_{8}=0, a_{2}=0\right\},\\
&s_{3}=\left\{a_{1}=0, a_{8}=0, a_{9}=0\right\},\\
&s_{4}=\left\{a_{8}=0, a_{9}=1, a_{2}=0\right\}.\label{e131}
\end{align}
From those sets, only the last set, \eqref{e131}, gives rise to a matrix ${\grL^{\gra}}_{\grb}$ with non-zero determinant
\begin{align}
{\grL^{\gra}}_{\grb}=\begin{pmatrix}
a_{1} & 0 & a_{3}\\
0 & a_{5} & a_{6}\\
0 & 0 & 1
\end{pmatrix}.\label{e132}
\end{align}
Finally, we may use parameters $b_{i}$, $(i=1,2,3,4)$ with $a_{1}=e^{b_{1}}, a_{3}=b_{2}, a_{5}=e^{b_{3}}, a_{6}=b_{4}$ such that
\begin{align}
{\grL^{\gra}}_{\grb}=\begin{pmatrix}
e^{b_{1}} & 0 & b_{2}\\
0 & e^{b_{3}} & b_{4}\\
0 & 0 & 1
\end{pmatrix},\label{e137}
\end{align}
(which is exactly the form presented in Table $1$), and for $b_{i}=0, (i=1,2,3,4)$ becomes the identity matrix.

\section{Einstein's equations under the assumption $\grg_{\grm\grn}=a(t)^{2}m_{\grm\grn}$}

Let us first start with the expressions of the Ricci tensor and the Ricci scalar. Note, that $\grm,\grn$,.. are triad indices, taking values from $1$ to $3$. The Ricci tensor depends on the structure constants of the algebra, the spatial metric and it's inverse as we can see from \eqref{e21}. The dependence of the metric and it's inverse comes always in pairs meaning $\grg^{\grm\grn}$$\grg_{\grs\grl}$, which implies that the scale factor $a(t)^{2}$ will cancel out resulting $R_{\grm\grn}(t)=\tilde{R}_{\grm\grn}$. Thus, for the Ricci scalar the following can be deduced $R(t)=\grg^{\grm\grn}(t)R_{\grm\grn}(t)=\frac{1}{a(t)^{2}}m^{\grm\grn}\tilde{R}_{\grm\grn}\Rightarrow R(t)=\frac{1}{a(t)^{2}}\tilde{R}$. Let us proceed with the extrinsic curvature tensor which can be calculated to be $K_{\grm\grn}(t)=-a(t)\dot{a}(t)m_{\grm\grn}$, while the corresponding scalar $K(t)=-3\frac{\dot{a}(t)}{a(t)}=-3H$. By use of these expressions in the equations \eqref{e17}, \eqref{e18}, \eqref{e19} and the choices $N_{\grm}=0, N=1$ the result is
\begin{align}
&\eqref{e17}\rightarrow H^{2}=\grk\frac{\grr^{(tot)}}{3}-\frac{\tilde{R}}{6}\frac{1}{a^{2}},\label{e209}\\
&\eqref{e18}\rightarrow q_{\grm}^{(tot)}=0,\\
&\eqref{e19}\rightarrow \frac{\ddot{a}}{a}m_{\grm\grn}=-2H^{2}m_{\grm\grn}+\frac{\grk}{2}\left(\grr^{(tot)}-P^{(tot)}\right)m_{\grm\grn}+\frac{\grk}{a^{2}}\grp_{\grm\grn}^{(tot)}-\frac{1}{a^{2}}\tilde{R}_{\grm\grn}.\label{e211}
\end{align}
Replacing $H^{2}$ in \eqref{e211} from \eqref{e209} results
\begin{align}
\frac{\ddot{a}}{a}m_{\grm\grn}=-\frac{\grk}{6}\left(\grr^{(tot)}+3 P^{(tot)}\right)m_{\grm\grn}+\frac{\grk}{a^{2}}\grp_{\grm\grn}^{(tot)}-\frac{1}{a^{2}}\left(\tilde{R}_{\grm\grn}-\frac{1}{3}\tilde{R}m_{\grm\grn}\right).\label{e212}
\end{align}
The trace of \eqref{e212} provide us with the only dynamical equation
\begin{align}
\frac{\ddot{a}}{a}=-\frac{\grk}{6}\left(\grr^{(tot)}+3 P^{(tot)}\right)\,\,\text{or}\,\, \dot{H}+H^{2}=-\frac{\grk}{6}\left(\grr^{(tot)}+3 P^{(tot)}\right),
\end{align}
which when is used back in \eqref{e212} constraints the traceless anisotropic pressure tensor to be 
\begin{align}
\grp_{\grm\grn}^{(tot)}=\frac{1}{\grk}\left(\tilde{R}_{\grm\grn}-\frac{1}{3}\tilde{R}m_{\grm\grn}\right).
\end{align}

\section{Proof of the statement that $\partial_{i}\grf(t,x)=\grf_{\gra}\grs^{a}_{i}(x)$, $\grf_{\gra}C^{\gra}_{\grb\grl}=0$}

For this section, the Greek coordinate-basis indices $\grm,\grn$,.. run from $0$ to $3$, the capital Latin triad indices $I,J,$.. run from $1$ to $3$ and finally the Latin coordinate-basis indices $i,j,$.. from $1$ to $3$. The energy momentum tensor $T^{(\grf)}_{\grm\grn}$ and it's trace $T^{(\grf)}$ for a free scalar field are
\begin{align}
T^{(\grf)}_{\grm\grn}=M\left(\partial_{\grm}\grf\partial_{\grn}\grf-\frac{1}{2}g_{\grm\grn}\partial_{\grs}\grf\partial^{\grs}\grf\right),\,\,
T^{(\grf)}=-M\frac{d-1}{2}\partial_{\grs}\grf\partial^{\grs}\grf,
\end{align}
where $d$ the dimension of the hypersurfaces. 

If the spacetime admits a Killing field, then the following relations hold
\begin{align}
{\cal{L}}_{\grj}g_{\grm\grn}=0\Rightarrow {\cal{L}}_{\grj}G_{\grm\grn}=0\Rightarrow{\cal{L}}_{\grj}T^{(tot)}_{\grm\grn}=0,
\end{align}
where the Einstein's equations were used from the second to the third step and $T^{(tot)}_{\grm\grn}$ the total energy momentum tensor. In the case that we study in section 4, the $T^{(tot)}_{\grm\grn}$ was decomposed into the energy momentum tensor of a sum of perfect fluids $T_{\grm\grn}^{(pf)}$ and $T_{\grm\grn}^{(\grf)}$. The equation ${\cal{L}}_{\grj}T^{(pf)}_{\grm\grn}=0$ is identically satisfied for the symmetry group of Bianchi Types, thus we are left with
\begin{align}
{\cal{L}}_{\grj}T^{(\grf)}_{\grm\grn}=0\Rightarrow
{\cal{L}}_{\grj}\left(\partial_{\grm}\grf\right)\partial_{\grn}\grf+\partial_{\grm}\grf{\cal{L}}_{\grj}\left(\partial_{\grn}\grf\right)-g_{\grm\grn}\partial^{\grs}\grf{\cal{L}}_{\grj}\left(\partial_{\grs}\grf\right)=0.\label{e202}
\end{align}  
We take the trace of \eqref{e202} and for $d\neq{1}$ we find 
\begin{align}
\partial^{\grs}\grf{\cal{L}}_{\grj}\left(\partial_{\grs}\grf\right)=0,\label{e203}
\end{align}
while using it back in \eqref{e202} results
\begin{align}
{\cal{L}}_{\grj}\left(\partial_{\grm}\grf\right)\partial_{\grn}\grf+\partial_{\grm}\grf{\cal{L}}_{\grj}\left(\partial_{\grn}\grf\right)=0.\label{e204}
\end{align}
The next step is to contract \eqref{e204} with $\partial^{\grs}\grf$ 
\begin{align}
\partial^{\grm}\grf{\cal{L}}_{\grj}\left(\partial_{\grm}\grf\right)\partial_{\grn}\grf+\partial^{\grm}\grf\partial_{\grm}\grf{\cal{L}}_{\grj}\left(\partial_{\grn}\grf\right)=0\xRightarrow{\eqref{e203}}
\partial^{\grm}\grf\partial_{\grm}\grf{\cal{L}}_{\grj}\left(\partial_{\grn}\grf\right)=0\Rightarrow\nonumber
\end{align}
\begin{align}
\partial^{\grm}\grf\partial_{\grm}\grf=0\,\,\, \text{or}\,\,\, {\cal{L}}_{\grj}\left(\partial_{\grn}\grf\right)=0.
\end{align}
There are two possibilities as we can see. In our case, $\partial_{t}\grf=0$ thus $\partial^{\grm}\grf\partial_{\grm}\grf=\partial^{i}\grf\partial_{i}\grf$, while the requirement of the spatial metric to be definite positive leads to $\partial^{i}\grf\partial_{i}\grf\neq{0}$. Hence, we are lead to the conclusion that
\begin{align}
{\cal{L}}_{\grj}\left(\partial_{\grn}\grf\right)=0.
\end{align}
This equation splits into temporal and spatial components. Considering also the set of Killing fields of the Bianchi Types we get
\begin{align}
&\grn=t,\,\,\, {\cal{L}}_{\grj_{I}}\left(\partial_{t}\grf\right)=0\Rightarrow\grj_{I}^{i}\partial_{i}\left(\partial_{t}\grf\right)=0\Rightarrow\partial_{t}\left(\partial_{i}\grf\right)=0,\label{e207}\\
&\grn=j,\,\,\, {\cal{L}}_{\grj_{I}}\left(\partial_{j}\grf\right)=0\Rightarrow\partial_{j}\grf(t,x)=\grf_{J}(t)\grs^{J}_{j}(x),\label{e208}
\end{align}
where the existence of the inverse of $\grj_{I}^{i}$ was used and the well behavior of $\grf$, $\partial_{i}\partial_{t}\grf=\partial_{t}\partial_{i}\grf$. By use of \eqref{e208} in \eqref{e207} the result is $\grf_{J}(t)=\grf_{J}$ since the inverse of $\grs^{J}_{j}$ exists. 

The final step is to take the spatial derivative of \eqref{e208}
\begin{align}
\partial_{l}\partial_{j}\grf=\grf_{J}\partial_{l}\grs^{J}_{j}
\end{align} 
interchange the indices $l,j$ and subtract while using the equation $\partial_{l}\partial_{j}\grf=\partial_{j}\partial_{l}\grf$
\begin{align}
\grf_{J}\left(\partial_{l}\grs^{J}_{j}-\partial_{j}\grs^{J}_{l}\right)=0\xRightarrow{\eqref{e15}}-\grf_{J}C^{J}_{IL}\grs^{I}_{l}\grs^{L}_{j}=0,
\end{align}
to find 
\begin{align}
\grf_{J}C^{J}_{IL}=0,
\end{align}
which completes the proof.

\section{The assumptions about $E_{i}(t,x),\,B_{ij}(t,x),\,J_{i}(t,x).$}

For this section, the convention for the indices are the same as in the previous one. As we have pointed out in Appendix C, when a metric admits a Killing field, the following relation holds
\begin{align}
{\cal{L}}_{\grj}g_{\grm\grn}=0\Rightarrow {\cal{L}}_{\grj}G_{\grm\grn}=0\Rightarrow{\cal{L}}_{\grj}T^{(tot)}_{\grm\grn}=0,
\end{align}
where $T_{\grm\grn}^{(tot)}$ the total energy momentum tensor. For our purposes, the total energy momentum tensor consists of a set of perfect fluids and electromagnetic field. In the specific case, which is also the case studied in this work, the perfect fluids satisfy identically the condition ${\cal{L}}_{\grj}T^{(fluids)}_{\grm\grn}=0$, then the only condition to be satisfied is ${\cal{L}}_{\grj}T^{(other)}_{\grm\grn}=0$, where $T_{\grm\grn}^{(em)}=\frac{1}{\grm_{0}}\left(F_{\grm\grs}{F_{\grn}}^{\grs}-\frac{1}{4}g_{\grm\grn}F\right)$ and $F=F_{\grs\grr}F^{\grs\grr}$. Note that in general, the symmetry is not ``inherited'' to the Faraday tensor:
\begin{align}
{\cal{L}}_{\grj}T_{\grm\grn}^{(em)}=0\neq>{\cal{L}}_{\grj}F_{\grm\grn}=0,
\end{align}
while the inverse is always true (given that the Lie derivative operator obeys the Leibnitz rule)
\begin{align}
{\cal{L}}_{\grj}F_{\grm\grn}=0\Rightarrow{\cal{L}}_{\grj}T_{\grm\grn}^{(em)}=0.
\end{align}
Further information about that can be found in the chapter $11$ of Part II of the book \cite{stephani_kramer_maccallum_hoenselaers_herlt_2003}. Let us express the Faraday tensor in the coordinate basis
\begin{align}
&\mathbf{F}=F_{\grm\grn}dx^{\grm}\otimes dx^{\grn}=F_{0i}dt\otimes dx^{i}+F_{i0}dx^{i}\otimes dt+F_{ij}dx^{i}\otimes dx^{j}\Rightarrow\nonumber\\
&\mathbf{F}=F_{0i}dt\wedge dx^{i}+F_{ij}dx^{i}\wedge dx^{j}.
\end{align} 
We introduce the electric and magnetic field as
\begin{align}\label{FD}
\mathbf{F}=-E_{i}(t,x)dt\wedge dx^{i}+B_{ij}(t,x)dx^{i}\wedge dx^{j}.
\end{align}
where $B_{ij}(t,x)=-B_{ji}(t,x)$. Let us perform a basis transformation, $dx^{i}=\grs^{i}_{\gra}(x)\grs^{\grs}$ where $\grs^{\gra}$ the one-form basis of Bianchi Types. The \eqref{FD} is now written as follows:
\begin{align}\label{FD1}
\mathbf{F}=-E_{i}(t,x)\grs^{i}_{\gra}(x)dt\wedge \grs^{\gra}+B_{ij}(t,x)\grs^{i}_{\gra}(x)\grs^{j}_{\grb}(x)\grs^{\gra}\wedge\grs^{b}.
\end{align}
All the symmetries of the spacetime (e.g. the Bianchi Types) will be inherited to the Faraday tensor, if and only if
\begin{align}
&E_{i}(t,x)\grs^{i}_{\gra}(x)=E_{\gra}(t)\Rightarrow\nonumber\\
&E_{i}(t,x)=E_{\gra}(t)\grs^{\gra}_{i}(x),\label{ED1}\\
&B_{ij}(t,x)\grs^{i}_{\gra}(x)\grs^{j}_{\grb}(x)=B_{\gra\grb}(t)\Rightarrow\nonumber\\
&B_{ij}(t,x)=B_{\gra\grb}(t)\grs^{\gra}_{i}(x)\grs^{\grb}_{j}(x),\label{ED2}
\end{align}
where $\grs^{\gra}_{i}(x)$ the inverse of $\grs^{i}_{\gra}$. Therefore, the previous equations are mere assumptions, since we might as well had
\begin{align}
&E_{i}(t,x)\grs^{i}_{\gra}(x)=E_{\gra}(t,x),\\
&B_{ij}(t,x)\grs^{i}_{\gra}(x)\grs^{j}_{\grb}(x)=B_{\gra\grb}(t,x),
\end{align}
in which case, some of the symmetries would have been inherited due to the relation ${\cal{L}}_{\grj}T_{\grm\grn}^{(em)}=0$, but not all of them. The same line of thought holds for the components of the current density $J_{\gri}(t,x)$. Finally, note that if \eqref{ED1},\eqref{ED2}, did not hold, then the equations (\textbf{CMFE}), (\textbf{MFE}), (\textbf{CCFE}) of the sub-section $2.4$ could not be written in the form they are.

\end{appendices}

\newpage

\acknowledgments{\begin{figure}[h!]
\centering
  \begin{subfigure}[h]{0.265\linewidth}
    \includegraphics[width=\linewidth]{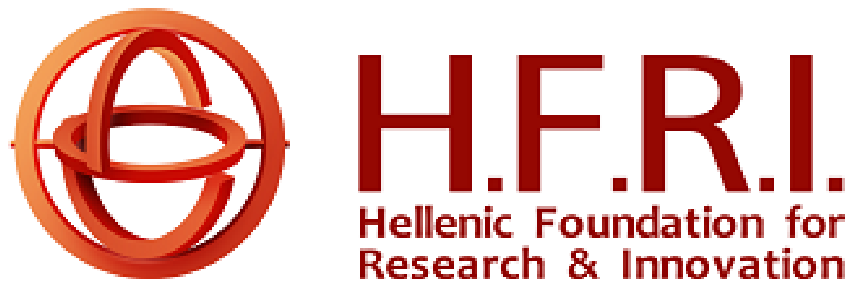}
  \end{subfigure}
  \begin{subfigure}[h]{0.2\linewidth}
    \includegraphics[width=\linewidth]{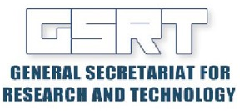}
  \end{subfigure}
\end{figure}

The research work was supported by the Hellenic Foundation for Research and Innovation (HFRI) and the General Secretariat for Research and Technology (GSRT), under the (HFRI) PhD Fellowship grant (GA.no.14501).

\bibliographystyle{unsrt}

\end{document}